\documentclass[a4paper,12pt]{article}
\usepackage{jheppub,esint,shuffle,psfrag}
 \usepackage[utf8]{inputenc}
 
\usepackage{tikz}
\usetikzlibrary{arrows.meta,decorations.markings}

\usepackage{hyperref}
\usepackage{esint} 
\usepackage{breqn}
\usepackage{bm}
\usepackage{caption}
\usepackage{amssymb}
\usepackage{placeins}\input

\def \tr {\mathop{\rm tr}\nolimits}
\def \Im {\mathop{\rm Im}\nolimits}
\def \Re {\mathop{\rm Re}\nolimits}
\def \res{\mathop{\rm res}\nolimits}
\newcommand\lr[1]{{\left({#1}\right)}}
\newcommand \widebar [1] {\overline{#1}}

\newcommand \vev [1] {\langle{#1}\rangle}
\newcommand \VEV [1] {\left\langle{#1}\right\rangle}

\newcommand\re[1]{(\ref{#1})}
\def \qqquad {\qquad\quad}
\def \qqqquad {\qquad\qquad}

\newcommand{\ft}[2]{{\textstyle\frac{#1}{#2}}}
\usepackage{comment}

\def\numberbysection{\@addtoreset{equation}{section}
                     \def\theequation{\thesection.\arabic{equation}}}

\begin{document}

\vspace*{0cm }

\author{Bercel Boldis$^{a,b}$, Gregory P. Korchemsky$^{c}$  and  Alessandro Testa$^{c}$  }
\affiliation{
$\null$
$^a$Department of Theoretical Physics, Institute of Physics, Budapest University of Technology and Economics M\H{u}egyetem rkp. 3., 1111 Budapest, Hungary
\\
$\null$ 
$^b$HUN-REN Wigner Research Centre for Physics, Konkoly-Thege Miklos ut 29-33, 1121 Budapest, Hungary
\\		
$\null$
$^c${Institut de Physique Th\'eorique\footnote{Unit\'e Mixte de Recherche 3681 du CNRS}, Universit\'e Paris Saclay, CNRS,  91191 Gif-sur-Yvette, France}   
}
\title{Bootstrapping ABJM theory}
\abstract{\small
Supersymmetric localization reduces the computation of protected observables in ABJM theory to finite-dimensional matrix integrals. Building on the techniques introduced in \href{https://arxiv.org/pdf/2512.02119}{arXiv:2512.02119}, we develop a bootstrap framework for the systematic calculation of instanton corrections to the free energy and to supersymmetric Wilson loops. Exploiting exact functional relations and consistency conditions satisfied by grand-canonical observables, in the Fermi-gas formulation of the ABJM matrix model, we provide analytic derivations of several relations for the free energy that were previously known only conjecturally, either from refined topological string theory or from high-precision numerical studies. We apply the same framework to determine the nonperturbative corrections to $1/2$ and $1/6$ BPS Wilson loops, elucidating their qualitative differences and uncovering novel structural features of the instanton effects. These results further highlight the intricate nonperturbative structure and network of dualities underlying ABJM theory.
}

\maketitle

\section{Introduction}

Three--dimensional $\mathcal{N}=6$ superconformal Chern--Simons--matter theory with gauge group $U(N)_k \times U(N)_{-k}$, known as ABJM theory~\cite{Aharony:2008ug,Aharony:2008gk}, provides a powerful framework for obtaining exact results in strongly coupled quantum field theory and holography. It describes the low-energy world-volume dynamics of coincident M2-branes and represents the best-understood example of the AdS$_4$/CFT$_3$ correspondence. In the large-$N$ limit and for fixed Chern--Simons level $k$, the ABJM theory is dual to an eleven-dimensional  M-theory on $AdS_4 \times S^7/\mathbb{Z}_k$ which, for large $k$ and fixed 't~Hooft coupling $\lambda = N/k$, reduces to type-IIA string theory on $AdS_4 \times \mathbb{CP}^3$. 
 
 A major advance in the study of ABJM theories was achieved through the application of supersymmetric localization~\cite{Pestun:2016zxk,Kapustin:2009kz}, which  reduces the partition function of ABJM theory on a round three-sphere, as well as the vacuum expectation values of multi-wound supersymmetric Wilson loops, to finite-dimensional matrix integrals. These expressions are exact and, in principle, encode the full nonperturbative dynamics of the theory at finite $N$ and Chern--Simons level $k$. 
 
The free energy on $S^3$ and the expectation values of circular $1/6$ BPS and $1/2$ BPS Wilson loops in ABJM theory have been investigated extensively over the past decade (for a review, see \cite{Hatsuda:2015gca,Marino:2016new}). In the large-$N$ limit, the free energy admits a decomposition into a perturbative contribution, organized as an asymptotic series in $1/\sqrt N$ and nonperturbative corrections that are exponentially suppressed  at large $N$. The perturbative part admits a remarkably compact representation in terms of an Airy function that captures the perturbative expansion to all orders. 

The nonperturbative corrections take the form of a double series of exponentially small terms of the form $e^{-2\pi \sqrt{N}h_{m,\ell}(k)}$, with  $h_{m,\ell}(k)=m \sqrt{2/k} + \ell \sqrt{k/2}$ and $m,\ell$ being nonnegative integers. In the M-theory description, these contributions arise from two kinds of M2-brane instantons, classified in \cite{Giombi:2023vzu,Beccaria:2023ujc} as type I and type II, and their bound states. 
Supersymmetric Wilson loops admit similar large-$N$ expansions: the perturbative sectors are again governed by Airy functions, while nonperturbative effects generate a rich instanton structure  that mirrors  the free energy, albeit with important qualitative differences.
 
Despite this impressive amount of results, their derivation remains largely  conjectural. The nonperturbative structure of the free energy and supersymmetric Wilson loops has been reconstructed using a combination of semiclassical expansions in the Fermi--gas formulation, conjectural relations to closed and open (refined) topological string theory on local $\mathbb{P}^1 \times \mathbb{P}^1$, high-precision numerical evaluations at finite $N$, and consistency conditions, such as pole-cancellation mechanism for special values of the level $k$. To date, there is no direct analytic derivation of the full instanton series starting from the localization matrix integrals. Providing such a derivation is the main goal  of the present work.

The Fermi--gas reformulation of ABJM matrix model  \cite{Marino:2011eh} provides a powerful framework for studying ABJM theory in the large-$N$ limit at fixed $k$.
 In this approach, the partition function is recast as that of a one-dimensional ideal quantum gas with $N$ particles and Planck constant $\hbar=2\pi k$.
This leads to an exact representation of the canonical partition function in terms of the grand potential $J(\mu,k)$ of the Fermi--gas ensemble with chemical potential $\mu$ \cite{Hatsuda:2012dt}
\begin{align}\label{pre-Airy}
Z(N,k) = \int_{-i\infty}^{i\infty} \frac{d\mu}{2\pi i}\, e^{J(\mu,k)-N\mu}\,.
\end{align} 
In the large-$N$ limit, the integral in \re{pre-Airy} receives a dominant contribution from a saddle point located at large $\mu=O(\sqrt N)$. 

At large chemical potential, the grand potential admits the expansion
\begin{align}\label{J-ans}
J(\mu,k)= C(k)\,\mu^{3} + B(k)\,\mu + A(k)+\sum_{m+\ell\ge 1}  f_{m,\ell}(\mu) \,e^{-({4m\over k}+2\ell)\mu}\,,
\end{align}
where $m$ and $\ell$ are nonnegative integers.
The first three terms in \re{J-ans} encode the perturbative contribution, while the remaining terms represent genuinely nonperturbative effects. The latter are governed by the coefficient functions $f_{m,\ell}(\mu)$, which depend on the Chern-Simons level $k$ and are polynomial in $\mu$.

In the holographic descriptions of ABJM theory, the different terms in the sum \re{J-ans} have the following interpretation. 
In the type-IIA string theory 
on ${\rm AdS}_4 \times \mathbb{CP}^3$, the non-perturbative contributions in the sum \re{J-ans} with $(m\neq 0,\, \ell =0)$ and $(m=0,\, \ell \neq 0)$  are identified with the world-sheet and membrane instantons, respectively. The former arise from fundamental strings wrapping the holomorphic cycle $\mathbb{CP}^1 \subset \mathbb{CP}^3$ \cite{Cagnazzo:2009zh,Drukker:2011zy}, whereas the latter are due to D2-branes wrapping the Lagrangian submanifold $\mathbb{RP}^3 \subset \mathbb{CP}^3$ \cite{Drukker:2011zy}. Moreover, the mixed contributions with $\ell,m\neq 0$ are interpreted as bound states of world-sheet and membrane instantons.  Upon lifting to M-theory, 
these two classes of instantons are unified in terms of M2-branes wrapping  three-cycles in $S^7/\mathbb Z_k$, which descend to $\mathbb{RP}^3$ and $\mathbb{CP}^1$ in the type-IIA description.
Consequently, a generic term in \re{Jnp} corresponds to a composite M2-brane configuration wrapping the two types of cycles $\ell$ and $m$ times, respectively.
 
The reformulation \re{pre-Airy} enables a systematic semiclassical WKB expansion of the partition function in the small-$k$ regime \cite{Marino:2011eh}. This expansion provides direct access to the perturbative part of the grand potential \re{J-ans} and captures 
membrane instanton contributions to \re{J-ans}.
By contrast, world-sheet instantons are not accessible by the semiclassical analysis, although they survive in the ’t~Hooft limit. Their structure was derived from the duality of the ABJM matrix model in the 't Hooft limit and the topological string theory on local $\mathbb{P}^1 \times \mathbb{P}^1$ \cite{Marino:2009jd, Drukker:2010nc} and confirmed by high-precision numerical analyses \cite{Hatsuda:2012dt}.

The situation becomes even more subtle for supersymmetric Wilson loops preserving $1/6$ and $1/2$ of the total $24$ superconformal symmetries \cite{Drukker:2008zx,Chen:2008bp,Drukker:2009hy,Drukker:2019bev}. Importantly, these two Wilson loops, when defined in the fundamental representation of the gauge group, are not independent but are related as \cite{Klemm:2012ii}
\begin{align}\label{W-half}
W_n^{1/2}=W_n^{1/6}-(-1)^n\,\overline{W_n^{1/6}}\,,
\end{align}
where the non-negative integer $n$ denotes the winding number and the bar denotes complex conjugation. As a consequence, it is sufficient to compute $1/6$ BPS Wilson loop. 

In the Fermi-gas formulation, the $n-$wound $1/6$ BPS Wilson loop $W_n=W_n^{1/6}$ is expressed as~\cite{Okuyama:2016deu}
\begin{align}\label{W-W}
W_n(N,k)=\int_{-i\infty}^{i\infty}\frac{d\mu}{2\pi i}\,\mathsf W^{1/6}_n(\mu,k)\,e^{J(\mu,k)-N\mu}\,.
\end{align} 
The function $\mathsf W^{1/6}_n(\mu,k)$ is obtained by averaging a specific single-particle operator in the grand canonical ensemble. At large $\mu$, it admits an expansion analogous to \re{J-ans}
\begin{align}\label{W-ans} 
 \mathsf W^{1/6}_n(\mu,k)={i^n e^{2n\mu/k}\over k\, \sin({2\pi n\over k})}\bigg[\mathsf w_{0}(\mu)+\sum_{m+\ell\ge 1}  \mathsf w_{m,\ell}(\mu) \,e^{-({4m\over k}+2\ell)\mu}\bigg]\,,
\end{align}
where the first term inside the brackets corresponds to the perturbative contribution and the different nonperturbative terms in the sum have the same interpretation as in \re{J-ans}. 

The exponential factor on the right-hand side of \re{W-ans} correctly reproduces the expected behavior of $1/2$ BPS Wilson in the M-theory regime \cite{Klemm:2012ii,Giombi:2023vzu}. Substituting the ansatz \re{W-ans} into \re{W-W} and performing the integration, one obtains an exact representation of $W_n(N,k)$ in terms of an infinite sum of Airy functions and their derivatives \cite{Okuyama:2016deu}.

The semiclassical approximation can be used to extract the leading
$O(\mu)$ behavior of the perturbative contribution in \re{W-ans}, but it fails already at the
subleading $O(\mu^0)$ order \cite{Klemm:2012ii}. The perturbative contribution $\mathsf w_{0}(\mu)$ was conjectured in \cite{Okuyama:2016deu} by numerical analysis and derived only recently in \cite{Boldis:2025yll}.

Furthermore, the structure of instanton
corrections to \re{W-ans} is substantially more intricate than that of
the grand potential \re{J-ans}.
For the $1/2$~BPS Wilson loop, closed-form expressions for
$W^{1/2}_n(N,k)$ were originally derived in \cite{Grassi:2013qva}  by invoking 
relations to open topological string amplitudes, leading to formulas
expressed in terms of Ooguri–Vafa (OV) invariants~\cite{Ooguri:1999bv}. By contrast, the
$1/6$~BPS Wilson loop does not admit a known topological string
interpretation, and its nonperturbative structure has so far been
reconstructed mainly from high-precision numerical data \cite{Okuyama:2016deu}. In both
cases, a first-principles analytic derivation of the full instanton
expansion directly from the localization matrix model is still
lacking.

These obstacles motivate the search for a framework that determines
the free energy and Wilson loop expectation values directly within
the ABJM matrix model, without recourse to dual descriptions or
numerical fitting. In this work, we propose such a framework, building
on the techniques developed in \cite{Boldis:2025yll}. Our approach exploits exact
functional relations and consistency conditions satisfied by the
grand canonical partition function and by the generating functions of
supersymmetric Wilson loops in the Fermi--gas formalism. When combined
with the ansatz \re{J-ans} and \re{W-ans}, this framework allows for
a systematic and fully analytic determination of nonperturbative
contributions to both quantities. As a result, we provide
first-principle derivations of several results for the free energy
and the $1/2$~BPS Wilson loop that were previously known only at a
conjectural level, and obtain new results
for $1/6$~BPS Wilson loop.

The paper is organized as follows. In Section~\ref{sec2}, we review the Fermi--gas formulation of the ABJM matrix model and summarize the known results for the perturbative and nonperturbative corrections to the partition function and supersymmetric Wilson loops in ABJM theory.

In Section~\ref{sec:fromFermitoTracy}, we present the technique for computing ABJM matrix integrals developed in~\cite{Boldis:2025yll}. The method is based on an operator reformulation of the partition function and supersymmetric Wilson loops using the one-particle density operator of the Fermi gas. Exploiting its analytic and algebraic properties, we express the Wilson loop $W_n(N,k)$ through a set of auxiliary functions $f_{m,\ell}(\mu)$ that encode the nonperturbative contributions to the grand potential~\re{J-ans}.

Substituting the  ansatz~\re{W-ans} into the integral representation~\re{W-W} yields an alternative expression for $W_n(N,k)$ involving a second set of functions $\mathsf{w}_{m,\ell}(\mu)$, which parametrize the nonperturbative corrections to the Wilson loop \re{W-ans}. Requiring consistency between these two representations leads to a closed system of functional constraints. Solving these constraints in Section~\ref{sec:4}, we bootstrap the functions $f_{m,\ell}(\mu)$ and $\mathsf{w}_{m,\ell}(\mu)$, thereby determining the nonperturbative contributions to the supersymmetric Wilson loops.
We conclude in Section~\ref{sec:5} with a brief discussion of the results and their implications. Technical details are collected in the appendices.

\section{From the Fermi gas to the refined topological string}
\label{sec2}

In this section, we briefly review the previously established results for the partition function of ABJM theory on the three-sphere $S^3$ and for supersymmetric circular Wilson loops in this model (see, e.g., \cite{Hatsuda:2015gca,Marino:2016new} for reviews). 

\subsection{Fermi gas formalism} 
 
Supersymmetric localization yields the following representation for the ABJM partition function and the expectation value of the $1/6$ BPS Wilson loop in the fundamental representation  
\begin{align}\label{Z-W} 
Z (N,k) = {1\over N} W_0(N,k)\,,  
\qqqquad
W_n^{1/6}(N,k)= {W_n(N,k)\over W_0(N,k)}\,.
\end{align}
The function  $W_n(N,k)$ depends on  the Chern--Simons level~$k$, the rank of the gauge group~$N$ and the winding number~$n$. It is given by a finite-dimensional matrix integral \cite{Kapustin:2009kz}
\begin{align}
\label{Z}
W_n(N,k) &= \frac{1}{(N!)^2} \int_{-\infty}^\infty \prod_{i=1}^N {d\mu_i d\nu_i\over (2\pi)^2} 
{\prod_{i<j} \left[2 \sinh \left( \frac{\mu_i - \mu_j}{2} \right) \right]^2 \left[2 \sinh \left( \frac{\nu_i - \nu_j}{2} \right) \right]^2 \over \prod_{i,j} \left[2 \cosh \left( \frac{\mu_i - \nu_j}{2} \right) \right]^2}
e^{ \frac{ik}{4\pi} \sum_{i=1}^N (\mu_i^2 - \nu_i^2)}\sum_{i=1}^N e^{n \mu_i} \,.
\end{align}
The last factor on the  right-hand side of \re{Z} grows exponentially for large values of~$\mu_i$ and the convergence of the integral is guaranteed  only if the Chern--Simons level satisfies the relation $k > 2n$.

A powerful approach to computing \re{Z} relies on the observation that the partition function $Z(N,k)$ can be recast into the partition function of a one-dimensional ideal Fermi gas of $N$ particles~\cite{Marino:2011eh}
\begin{align}\label{Z-perm}
Z(N,k) = {1\over N!} \sum_{\sigma\in S_N} (-1)^{\epsilon(\sigma)}\int{dx_1 \dots dx_N}\prod_{i=1}^N \rho(x_i,x_{\sigma(i)})\,.
\end{align}
Here the sum goes over the permutations of $N$ particles  and the canonical density matrix is given by
\begin{align}\label{rho-fun}
\rho(x,y)={1\over 8\pi k  \cosh(\frac{x}2) \cosh({x-y\over 2k})}\,.
\end{align}
This function can be interpreted as the kernel of the integral operator $\bm{\rho}$ (see (\ref{rho-oper}) below), which plays a central role in our analysis. The analogous expression for the Wilson loop in terms of the density matrix is given below in \re{det2}.

It is advantageous to combine the functions \re{Z-W} for different $N$ and define the generating functions
\begin{subequations}
\begin{align}\label{Xi}
{}& \Xi(\mu,k) = 1+ \sum_{N\ge 1} e^{N\mu} Z(N,k)\,,
\\\label{cal-W}
{}& \mathcal W_n(\mu,k) = {1\over \Xi(\mu,k)}\sum_{N\ge 1} e^{N\mu} W_n(N,k)\,,
\end{align} 
\end{subequations}
where $z=e^\mu$ is the fugacity parameter. Defined in this way, the function $\Xi(\mu,k)$ can be interpreted as the grand canonical partition function of the Fermi gas with the chemical potential $\mu$. In a similar manner, 
the function $\mathcal W_n(\mu,k)$ describes the expectation value of the holonomy $\tr(U^n)=\sum_{i=1}^N e^{n \mu_i}$ in the grand canonical ensemble. 

By inverting the relations \re{Xi}, the partition function and the (un-normalized) Wilson loop can be found as
\begin{subequations}
\begin{align}\label{Z,W}
{}& Z(N,k) = \int_{-i\pi}^{i\pi} {d\mu\over 2\pi i} e^{-N\mu} \, \Xi(\mu,k)\,, 
\\\label{Z,W2}
{}& W_n(N,k) = \int_{-i\pi}^{i\pi} {d\mu\over 2\pi i} e^{-N\mu} \, \Xi(\mu,k) \mathcal W_n(\mu,k)\,.
\end{align}
\end{subequations}
For small values of $N$, the matrix integral \re{Z} can be evaluated explicitly for arbitrary $n$ and $k$, yielding closed-form expressions for the functions \re{Z-W}, see e.g. \cite{Hatsuda:2012hm,Putrov:2012zi}. This regime corresponds to the dilute limit of the Fermi gas and is captured by the large negative-$\mu$ regime of the generating function \re{Xi}.
Conversely, computing the matrix integral \re{Z} in the large-$N$ regime is substantially more challenging.  

At large $N$, the observables \re{Z-W} exhibit distinct properties depending on the scaling of the Chern--Simons level $k$. In the type-IIA string theory regime, corresponding to the 't Hooft limit $N,k\to\infty$ with their ratio $\lambda=N/k$ kept fixed, the matrix integral \re{Z} 
admits a systematic topological expansion in powers of $1/N^2$.
 In the M-theory regime, for $N\to\infty$ with fixed $k$, the Fermi gas approach provides a powerful framework for computing the observables \eqref{Z-W}, capturing non-perturbative effects that are typically inaccessible via standard topological expansions. Within this approach, the large-$N$ limit is governed by the thermodynamic behavior of the generating functions \re{Xi} at large positive $\mu$.
  
\subsection{The grand canonical partition function}
\label{sec:MGCP}

In the large-$N$ limit, the partition function $Z(N,k)$ admits the integral representation \re{pre-Airy}. In contrast to \re{Z,W}, the integration contour in \eqref{pre-Airy} runs along the entire imaginary axis. By decomposing the contour in \eqref{pre-Airy} into an infinite sequence of segments of length $2\pi$ and performing an appropriate shift of the integration variable on each segment, one can recover \eqref{Z,W} and identify the grand canonical partition function as \cite{Hatsuda:2012dt}
\begin{align}\label{Xi-conj}
\Xi(\mu,k) = \sum_{m=-\infty}^{\infty} e^{J(\mu + 2\pi i m, k)} \,.
\end{align}
As is manifest from its definition \re{Xi}, the function $\Xi(\mu,k)$ is invariant under shifts $\mu \to \mu + 2\pi i$. The representation \eqref{Xi-conj} makes this periodicity explicit.

The grand potential $J(\mu,k)$ has been determined at large $\mu$ and finite $k$ by first computing it in the small-$k$ limit and then extending the result to arbitrary $k$ through a set of conjectures supported by high-precision numerical analysis. In particular, the resulting expression for  $J(\mu,k)$ can be decomposed into  perturbative and nonperturbative contributions\  as follows \cite{Marino:2011eh}
\begin{align}\label{J-grand}
J(\mu, k) = J_{\text{pert}}(\mu, k) + J_{\text{np}}(\mu, k)\,.
\end{align}
The perturbative contribution is a cubic polynomial in $\mu$, with coefficients that depend on the Chern--Simons level $k$,
\begin{align}\label{J-pt}
J_{\text{pert}}(\mu, k) = C(k)\,\mu^{3} + B(k)\,\mu + A(k)\,.
\end{align}
These coefficients were determined by analyzing the partition function \re{Z-perm} in the small-$k$ regime, which it corresponds to the semiclassical description of the Fermi gas and can be systematically studied using a WKB expansion.
The first two coefficients in \eqref{J-pt} are given by \cite{Marino:2011eh}
\begin{align}\label{CB}
C(k) = \frac{2}{\pi^{2}k}\,, \qqqquad 
B(k) = \frac{1}{3k} + \frac{k}{24}\,.
\end{align}
The remaining coefficient $A(k)$ admits a WKB expansion as a formal power series in $k$. An all-orders expression for $A(k)$ was conjectured in \cite{Hanada:2012si}.

The nonperturbative contribution to \re{J-grand}  is given by an infinite series of terms that are exponentially suppressed at large $\mu$
\begin{align}\label{Jnp}
J_{\text{np}}(\mu, k) = 
\sum_{m+\ell\ge 1} 
f_{m,\ell}(\mu)\, 
e^{-\left(\frac{4m}{k} + 2\ell\right)\mu} \,,
\end{align}
where $m$ and $\ell$ are nonnegative integers and the coefficients $ f_{m,\ell}(\mu) $ are polynomials in $\mu$ \footnote{To be more precise, only the membrane sector, namely $f_{0,\ell}$, displays a polynomial dependence on $\mu$.}. 
 As mentioned in the Introduction, the various terms in \re{Jnp} correspond to different kinds of M2-brane instantons. Although the M-theory description is more appropriate in the regime of large $\mu$ and finite $k$, we follow the type IIA string theory nomenclature of \cite{Hatsuda:2015gca,Marino:2016new} and classify the contributions in \re{Jnp} as follows: worldsheet instantons $(m \ge 1,\ell=0)$, membrane instantons $(m=0,\ell \ge 1)$, and their bound states $(m \ge 1,\ell \ge 1)$. 
Accordingly, the sum in \eqref{Jnp} can be conveniently organized into three distinct classes of corrections \cite{Hatsuda:2012dt} 
\begin{align}\label{J-b}
J_{\text{np}}(\mu,k)=J_{\text{WS}}(\mu,k)+J_{\text{M2}}(\mu,k)+J_{\text{BS}}(\mu,k) \,.
\end{align}
In what follows we describe the properties of each function separately. 

The world-sheet function in \re{J-b} is given by series in $e^{-4\mu/k}$
\begin{align}
\label{JWS}
J_{\text{WS}}(\mu,k)=\sum_{m\ge 1} d_m(k)\, e^{-{4m\over k}\mu}\,,
\end{align}
where the coefficients $d_m\equiv f_{m,0}$ are independent of $\mu$. This function vanishes for $k\to 0$ and, therefore, it is not captured by semiclassical expansion of the Fermi gas. However, 
it survives in the 't Hooft limit, for  $N\to\infty$ with $\lambda=N/k$ kept fixed. In this regime, the integral in \re{pre-Airy} can be approximated by the saddle point at $\mu\sim \pi\sqrt{Nk/2}$, so that 
$e^{-{4m \mu/k}}\sim e^{-2m\sqrt{2\lambda}}$ approaches a finite value.  

Using this property, the coefficients in \re{JWS} were computed in  \cite{Hatsuda:2013gj} by exploiting the duality between the large-$N$ expansion of the ABJM partition function in the 't Hooft limit and the topological string theory on the Calabi-Yau manifold known as local $\mathbb{P}^1\times \mathbb{P}^1$ \cite{Marino:2009jd, Drukker:2010nc}. These coefficients are given by a linear combinations of diagonal integer valued Gopakumar-Vafa  invariants $n_g^d$ \cite{Gopakumar:1998ii,Gopakumar:1998jq}
\begin{equation}
    \label{fm0}
    d_m(k) = \sum_{g=0}^{\infty}\sum_{d|m} n_d^g \frac{(-1)^m d}{m}\left(2 \sin\frac{2\pi m}{ d k}\right)^{2g-2},
\end{equation} 
where the sum runs over divisors $d$ of $m$. The explicit expressions for $d_m(k)$ for $m=1,2$ are given by
\begin{align}\notag
\label{f1f2f3}
{}& d_1(k)=\frac{1}{\sin^2\left(\frac{2\pi}{k}\right)}\,,
\\[2mm]
{}& d_2(k)=-\frac{1}{2\sin^2\left(\frac{4\pi}{k}\right)}-\frac{1}{\sin^2\left(\frac{2\pi}{k}\right)} \,.
\end{align} 
Note that functions $d_m(k)$ develop double poles for rational and integer values of  $k$. 
 
The contribution of membrane instantons to (\ref{J-b}) is given by a series in $e^{-2\mu}$
\begin{align}
J_{\text{M2}}(\mu,k)=\sum_{\ell\ge 1} f_{0\ell}(\mu,k)\, e^{-2\ell\mu} \,.
\end{align}
This function vanishes in the 't Hooft limit but it survives in the M-theory regime. Furthermore, its behavior at small $k$ was originally derived in  \cite{Marino:2011eh} by  a WKB approximation of the Fermi gas which leads to
\begin{align}
\label{f0l}
f_{0,\ell}(\mu,k)=\mu^2 a_\ell(k) + \mu\, b_\ell(k)+c_\ell(k) \,.
\end{align}
The expansion coefficients $a_\ell(k), b_\ell(k)$ and $c_\ell(k)$ can be computed using the WKB expansion of the Fermi gas as a series in $k$ \cite{Calvo:2012du}. However, they develop poles at finite values of $k$ \cite{Hatsuda:2012dt}. 

The additional condition for the membrane coefficients in \re{f0l} arises from the requirement that the nonperturbative function \re{Jnp} should be finite for integer and rational $k$. 
For this to happen spurious poles of the world-sheet function \re{JWS} have to cancel against similar divergences of the two remaining functions in \re{J-b}. Being combined with the small $k$ expansion, this pole cancellation mechanism was used in \cite{Hatsuda:2012dt} to determine the 1-membrane instanton coefficients in \re{f0l}.
 
However, extending the same mechanism to the membrane functions \re{f0l} at large values of $\ell$ is not sufficient to uniquely determine the coefficients $a_\ell$, $b_\ell$ and $c_\ell$. The underlying reason is that the bound-state contribution to \re{J-b} takes the form
\begin{equation}
\label{JBS}
J_{\text{BS}}(\mu,k)
=
\sum_{m,\ell\neq 0}
f_{m,\ell}(k)\,
\mathrm{e}^{-\frac{4 m \mu}{k}-2 \ell \mu}\,,
\end{equation}
and involves a new set of functions $f_{m,\ell}(k)$ that develop spurious poles at integer values of $k$ as well. As the membrane wrapping number $\ell$ increases, the number of unknown coefficients contributing to the residues at these spurious poles grows rapidly. Consequently, the resulting pole-cancellation conditions become increasingly intricate and are no longer sufficient to fix the coefficients $a_\ell$, $b_\ell$, and $c_\ell$.

\subsection{Effective chemical potential}

It was conjectured in \cite{Hatsuda:2013gj} that the contribution of the bound states to \re{J-b} can be absorbed into a redefinition of the chemical potential $\mu$. The effective chemical potential is defined as
\begin{equation}
    \label{eq:mueff}
    \mu_{\mathrm{eff}}= \mu +\frac{1}{C(k)}\sum_{\ell=1}^\infty a_\ell(k) e^{-2\mu \ell}\,,
\end{equation}
where the parameter $C(k)$ and the coefficients $a_\ell(k)$ are introduced in \re{CB} and \re{f0l}. 

Being expressed in terms of $\mu_{\mathrm{eff}}$, the grand potential \re{J-grand} is expected to have the following form~\cite{Hatsuda:2013gj}
\begin{align}\label{J-eff}
J(\mu,k) = J_{\text{pert}}(\mu_{\rm eff},k)+ J_{\text{WS}}(\mu_{\rm eff},k)+\widetilde J_{\text{M2}}(\mu_{\rm eff},k) \, ,
\end{align} 
where the perturbative and world-sheet functions are the same as in \re{J-pt} and \re{JWS}, up to redefinition of their argument. Conversely, the membrane function in \re{J-eff} 
is given by 
\begin{align}\label{JM2}\notag
\widetilde J_{\text{M2}}(\mu_{\rm eff},k) {}& =\sum_{\ell\ge 1} \lr{\mu_{\rm eff}\,\tilde b_\ell(k)+\tilde c_\ell(k)}e^{-2\mu_{\rm eff}\ell} 
\\
{}& = \mu_{\rm eff}\widetilde J_b(\mu_{\rm eff},k) + \widetilde J_c(\mu_{\rm eff},k)\,,
\end{align} 
where the notation was introduced for $\widetilde J_b(\mu_{\rm eff},k)=\sum \tilde b_\ell(k) e^{-2\mu_{\rm eff}\ell}$ and similar for $\widetilde J_c(\mu_{\rm eff},k)$.

In contrast to \re{J-b}, the right-hand side of \re{J-eff} does not contain an explicit bound-state function involving mixed exponential terms of the form
$e^{-\left(  {4m}/{k} + 2\ell \right)\mu_{\rm eff}}$. The bound-state contribution present in \re{J-b} emerges implicitly from re-expanding (\ref{JM2})
in terms of the original chemical potential~$\mu$.

This observation implies that the bound-state coefficients appearing in \re{JBS} are not independent: they can be expressed entirely in terms of the world-sheet instanton coefficients $d_m(k)$ and the leading membrane instanton coefficients $a_\ell(k)$ defined in \re{JWS} and \re{f0l}, respectively. Likewise, the modified membrane coefficients $\tilde{b}_\ell(k)$ and $\tilde{c}_\ell(k)$ can be determined from the original membrane instanton data $a_\ell(k)$, $b_\ell(k)$, and $c_\ell(k)$. Explicit expressions for these relations can be found in~\cite{Hatsuda:2013yua}. 

Furthermore, it was conjectured in \cite{Hatsuda:2013gj} that there is a nontrivial relationship between the coefficients $\widetilde c_\ell(k)$ and $\widetilde b_\ell(k)$. This means that the grand potential depends on only two sets of independent coefficients $a_\ell(k)$ and $\widetilde b_\ell(k)$ which, for the first few values of $\ell$, are given by \cite{Hatsuda:2013gj,Calvo:2012du}
\begin{align}\notag\label{AB}
{}&a_1(k) =-{4\over\pi^2 k}\cos\Big({\pi k\over 2}\Big)\,, &&  \widetilde b_1(k) = {2\over\pi} \cot\Big({\pi k\over 2}\Big)\cos\Big({\pi k\over 2}\Big)\,,
\\[2mm]
{}&a_2(k)=-{2\over\pi^2k}(4+5\cos(\pi k))\,,  && \widetilde b_2(k) = {2\over\pi} \cot (\pi k)(4+5\cos(\pi k))\,.
\end{align}
Similar to the membrane coefficients in \re{f0l}, the coefficients $\widetilde b_\ell(k)$ develop spurious poles at integer and rational $k$. As in the previous case, these poles cancel in the grand potential \re{J-eff} against the spurious poles of the world-sheet function $J_{\text{WS}}(\mu_{\rm eff},k)$.

An efficient method for the exact computation of the coefficients $a_\ell(k)$ and $\widetilde b_\ell(k)$ for arbitrary values of the Chern--Simons level $k$ was developed in~\cite{Hatsuda:2013oxa}. This approach is based on the conjecture that these coefficients are governed by the refined topological string on local $\mathbb{P}^1 \times \mathbb{P}^1$ in the Nekrasov--Shatashvili (NS) limit. More precisely, the coefficients $a_\ell(k)$ and $\widetilde b_\ell(k)$ can be identified with quantum periods of the spectral curve associated with local $\mathbb{P}^1 \times \mathbb{P}^1$, thereby providing a powerful framework for determining membrane instanton corrections to arbitrary order in the wrapping number~$\ell$.

\subsection{Supersymmetric Wilson loops}

According to \re{Z-W} and \re{W-half}, the expectation values of multi-wound supersymmetric Wilson loops
are expressed in terms of the same function $W_n(N,k)$ defined in 
\re{Z}. 

In close analogy with \re{pre-Airy}, this function admits, in the
large-$N$ limit, the integral representation \re{W-W}.  
Substituting \re{W-W} into \re{Z-W}, we find that the
(un-normalized)  $1/2$ BPS Wilson loop admits the representation~\re{W-W}, with
the function $\mathsf W^{1/6}_n(\mu,k)$ replaced by 
\begin{align}
\mathsf W^{1/2}_n(\mu,k)=\mathsf W^{1/6}_n(\mu,k)-(-1)^n\,\widebar{\mathsf W^{1/6}_n(\mu,k)}\,,
\end{align}
where the bar denotes complex conjugation.

The function $\mathsf W^{1/6}_n(\mu,k)$  is related to the generating function $ \mathcal W_n(\mu,k)$ defined in \re{cal-W} as
\begin{align}\label{W-GC}
 \mathcal W_n(\mu,k) = {1\over\Xi(\mu,k)} \sum_{m=-\infty}^\infty \mathsf W^{1/6}_n(\mu+2\pi im,k)e^{J(\mu+2\pi i m,k)}\,.
\end{align}
Indeed, substituting this relation into \re{Z,W2} and changing the integration variable inside the sum over $m$  as $\mu\to \mu-2i\pi m$ we recover \re{W-W}.  As in the case of $\Xi(\mu,k)$, the sum in \re{W-GC} restores the invariance of $\mathcal W_n(\mu,k)$ under $\mu\to \mu+2i\pi$. 

In the Fermi gas approach, the relation \re{W-GC} is interpreted as defining the expectation value of $\mathsf W_n(\mu,k)$ in the grand canonical ensemble \cite{Klemm:2012ii}
\begin{align}\label{vev-GC}
 \mathcal W_n(\mu,k) = \vev{\mathsf W^{1/6}_n(\mu,k)}_{_{\rm GC}}\,,
\end{align}
where the average $\vev{\dots}_{_{\rm GC}}$ is defined in \re{vev-GC2} below (see also Appendix~\ref{appA}). 
The computation of supersymmetric Wilson loops at large $N$ reduces to finding the functions $\mathsf W^{1/6}_n(\mu,k)$ and $\mathsf W^{1/2}_n(\mu,k)$ in the large-$\mu$ regime. 

These functions admit the following representation \cite{Klemm:2012ii,Hatsuda:2013yua,Okuyama:2016deu}
\begin{align}\notag\label{WW-gen}
{}& \mathsf W^{1/6}_n(\mu,k)={i^n e^{2n\mu/k}\over k\, \sin({2\pi n\over k})}\left[\mathsf W_n^{\rm pert}(\mu,k) +\mathsf W_n^{\rm np}(\mu,k) \right],
\\
{}& \mathsf W^{1/2}_n(\mu,k)=-{2i^{n-1} e^{2n\mu/k}\over k\, \sin({2\pi n\over k})}\Im \left[\mathsf W_n^{\rm pert}(\mu,k) +\mathsf W_n^{\rm np}(\mu,k) \right],
\end{align} 
where the two terms inside the brackets correspond to the perturbative and non-perturbative contributions, respectively. 
The exponential factor in \re{WW-gen} ensures that, upon replacing $\mu$ with its saddle point value $\mu \sim \pi\sqrt{Nk/2}$,  
the supersymmetric Wilson loops exhibit the leading large-$N$ behavior $\sim e^{n\pi \sqrt{2\lambda}}$, in agreement with the dual M-theory description \cite{Klemm:2012ii,Giombi:2023vzu}.

The general properties of the functions $\mathsf W_n^{\rm pert}(\mu,k)$ and $\mathsf W_n^{\rm np}(\mu,k)$ closely parallel those of the corresponding contributions to the grand potential~\re{J-grand}. An important distinction, however, is that these functions develop non-vanishing imaginary parts. The perturbative contribution to~\re{WW-gen} takes the form
\begin{align}\label{W-pert}
\mathsf W_n^{\rm pert}(\mu,k)
= \frac{\mu}{\pi}
- \sum_{j=1}^n \cot\!\left(\frac{2\pi j}{k}\right)
- \frac{i k}{4}\,.
\end{align}
The leading $O(\mu)$ term can be derived from the semiclassical description of the Wilson loop in the Fermi-gas formalism at small $k$~\cite{Klemm:2012ii}. By contrast, an attempt to compute the constant $O(\mu^0)$ term in \re{W-pert} reveals a breakdown of the semiclassical approximation. In particular, for $n>1$ the resulting value of the constant term differs from that obtained through high-precision numerical evaluation~\cite{Okuyama:2016deu}. This mismatch reflects a general limitation of the semiclassical approach in the computation of supersymmetric Wilson loops. The constant term in~\re{W-pert} was first conjectured on the basis of numerical analysis in~\cite{Okuyama:2016deu}, and was derived analytically only recently in~\cite{Boldis:2025yll}.

The nonperturbative contribution to \re{WW-gen} is given by the sum of world-sheet and membrane instantons
\begin{align}\label{W-np}
\mathsf W_n^{\rm np}(\mu,k)= \sum_{m+\ell\ge 1} \mathsf w_{m,\ell}(\mu)\, e^{-\left(\frac{4m}{k} + 2\ell\right)\mu} \,,
\end{align}
where the coefficients $\mathsf w_{m,\ell}(\mu)$ depend on the winding number $n$. The structure of the non-perturbative corrections~\re{W-np} remains far less understood than that of the corresponding contributions to the grand potential~\re{J-eff}. 

In a close analogy with \re{J-b}, the nonperturbative function \re{W-np} can be separated into the sum of contributions from the world-sheet instantons, the membrane instantons and their bound states
\begin{align}\label{W's}
\mathsf W_n^{\rm np}(\mu,k)=\mathsf W_n^{\rm WS}(\mu,k)+\mathsf W_n^{\rm M2}(\mu,k)+\mathsf W_n^{\rm BS}(\mu,k)\,.
\end{align}
The nonperturbative contribution to $1/2$ BPS Wilson loop \re{WW-gen} depends on the imaginary part of these functions. 

\paragraph{World-sheet instantons.}
For the $1/2$ BPS Wilson loop \re{WW-gen}, the contribution of the world-sheet instantons to $\Im \mathsf W_n^{\rm np}(\mu,k)$ can be determined by exploiting the relation of the observable 
to the open topological string on local $\mathbb{P}^1\times \mathbb{P}^1$. This leads to the following relation for the imaginary part of the world-sheet contribution \cite{Hatsuda:2013yua}
\begin{align}\label{WS-conj}
 \Im \mathsf W_n^{\rm WS}(\mu,k) ={k\over 4}  -{k\over 4} f(Q^n)\sum_{g,d\ge 0}  \sum_{\ell m=n} N_{\vec e_m,d}^g  
\lr{2\sin{2\pi\ell\over k}}^{2g-2} \lr{2\sin{2\pi n\over k}}^2 Q^{d\ell}\,,
\end{align}
where $Q=-e^{-4\mu/k}$, the integers $N_{\vec e_m,d}^g$ are related to the OV invariants and the second sum runs over multiples of $n$. Moreover, the notation was introduced for  
\begin{align}
f(Q)=1+2 Q+3 Q^2+10 Q^3+49 Q^4+288 Q^5+1892 Q^6 +O(Q^7)\,.
\end{align} 
The function $\Im \mathsf W_n^{\rm WS}(\mu,k)$ is given by series in $Q$ and it vanishes for $Q=0$.

For the $1/6$ BPS Wilson loop, a relation between world-sheet instantons and open topological string amplitudes, analogous to \re{WS-conj}, has not yet been established in the literature. As a result, the real part of the world-sheet non-perturbative contribution,
$\Re\,\mathsf W_n^{\rm WS}(\mu,k)$, remains largely undetermined. This function was investigated in~\cite{Okuyama:2016deu} using numerical analysis.

\paragraph{Membrane instantons.} 
For the $1/2$ BPS Wilson loop~\re{WW-gen}, the contributions of membrane instantons and their bound states to $\Im\,\mathsf W_n^{\rm np}(\mu,k)$ were studied numerically in~\cite{Hatsuda:2013yua}. Based on these results, it was conjectured in~\cite{Hatsuda:2013yua} that, in close analogy with the grand potential~\re{J-eff}, the membrane corrections can be incorporated by replacing the chemical potential $\mu$ with its effective counterpart 
$\mu_{\rm eff}$~\re{eq:mueff}. This leads to 
\begin{align}\label{W-1/2-conj}
\mathsf W^{1/2}_n(\mu,k)
= -\frac{2 i^{\,n-1} e^{2n\mu_{\rm eff}/k}}{k\, \sin\!\left(\frac{2\pi n}{k}\right)}
\left[
-\frac{k}{4}
+ \Im\,\mathsf W_n^{\rm WS}(\mu_{\rm eff},k)
\right].
\end{align}
In contrast to the structure of the grand potential~\re{J-eff}, the expression inside the brackets does not involve an explicit membrane function.
\footnote{We would like to emphasize that the membrane contribution arises implicitly upon re-expanding the Wilson loop in terms of the original chemical potential $\mu$. \label{foot}}
The conjecture~\re{W-1/2-conj} implies that the membrane-instanton and bound-state contributions to the $1/2$ BPS Wilson loop, defined in \re{WW-gen} and \re{W's} and
described by the functions $\Im\,\mathsf W_n^{\rm M2}(\mu,k)$ and $\Im\,\mathsf W_n^{\rm BS}(\mu,k)$, 
can be obtained by re-expanding the right-hand side of~\re{W-1/2-conj} in the chemical potential $\mu$. 

Numerical analyses revealed that the membrane-instanton contributions to the remaining functions, $\Re\,\mathsf W_n^{\rm M2}(\mu,k)$ and $\Re\,\mathsf W_n^{\rm BS}(\mu,k)$, exhibit a significantly more intricate structure~\cite{Okuyama:2016deu}. In contrast to \re{W-1/2-conj}, rewriting the $1/6$ BPS Wilson loop in terms of the effective chemical potential $\mu_{\rm eff}$ does not eliminate the contributions from membrane instantons, nor from their bound states with world-sheet instantons. Based on numerical fitting, a conjecture for the general structure of the membrane contributions to the $1/6$ BPS Wilson loop was proposed in~\cite{Okuyama:2016deu}. 

We emphasize that the results for the grand potential and supersymmetric Wilson loops presented in this section are based on a sequence of conjectures, motivated by the relation between ABJM theory and topological string theory and corroborated by dedicated numerical analyses. It remains highly desirable to develop an analytical framework capable of computing these quantities from first principles. Such an approach would provide a rigorous derivation -- and ultimately a formal proof -- of the conjectures discussed above.

In this paper, we take a step toward this goal. In the following section, we introduce a novel technique for evaluating the matrix integral~\re{Z}. We then apply this method to compute the grand potential and supersymmetric Wilson loops, and compare our results with the conjectured expressions discussed above.

\section{From the Fermi gas to Tracy-Widom}
\label{sec:fromFermitoTracy}

The starting point of our analysis is the expression for the generating functions~\re{Xi} and~\re{cal-W} in the Fermi-gas formulation. 

Upon replacing the functions $Z(N,k)$ and $W_n(N,k)$ by their integral representations \re{Z-W} and \re{Z}, we obtain the following  operator representation~\cite{Marino:2011eh,Klemm:2012ii,Okuyama:2016deu}
\begin{subequations}
\begin{align}\label{det1}
{}& \Xi(\mu,k)= \det\!\left(1 + z\, \boldsymbol{\rho}\right),
\\[0.2cm]\label{det2}
{}& \mathcal{W}_n(\mu,k)
= \tr\!\left(
\frac{z\, \boldsymbol{\rho}}{1 + z\, \boldsymbol{\rho}}\,
e^{\,n(\boldsymbol{x}+\boldsymbol{p})/k}
\right),
\end{align}
\end{subequations}
where $z=e^{\mu}$ is the fugacity parameter and $\boldsymbol{\rho}$ is the one-particle density operator of the Fermi gas.
This operator has a simple representation in terms of the standard quantum-mechanical position and momentum operators, $\boldsymbol{x}$ and $\boldsymbol{p}$, satisfying the canonical commutation relation
$[\boldsymbol{x},\boldsymbol{p}]=2\pi i k$,
\begin{align}\label{rho-oper}
\boldsymbol{\rho}
= \frac{1}{2\cosh(\boldsymbol{x}/2)}\,
\frac{1}{2\cosh(\boldsymbol{p}/2)}\,.
\end{align}
In the coordinate representation, the kernel of this operator $\rho(x,y) =\langle x | \boldsymbol{\rho} | y \rangle$ coincides with the density matrix \re{rho-fun}. 
 
Using the Baker-Campbell-Hausdorff formula  for $e^{n(\boldsymbol{x} + \boldsymbol{p})/k}$, the relation \re{det2} can be written in an equivalent form \cite{Okuyama:2016deu}
\begin{align}
\label{det3}
\mathcal{W}_n(\mu,k) = \int_{-\infty}^\infty dx\, e^{n( x+i\pi n)\over k}\VEV{x\left| \frac{z\, \boldsymbol{\rho}}{1 + z\, \boldsymbol{\rho}}\right|x+2\pi i n}.
\end{align} 
A notable feature of this representation is that the matrix element on the right-hand side involves an argument shifted into the complex plane by $2\pi in$. 

At large negative $\mu$, or equivalently small $z$, the function \re{det3} can be expanded in powers of $z$ with the coefficients given by convolution of the kernels \re{rho-fun}. In the following subsection, we derive an expression for $\mathcal{W}_n(\mu,k)$ that is valid for arbitrary $\mu$.

\subsection{Exact relations}

Following the approach of Tracy-Widom \cite{Tracy:1995ax}, we  introduce the following auxiliary function
\begin{align}\label{psi} 
\psi(x|z) {}&=\VEV{E\left|{1\over 1-z \bm\rho}\right|x} 
 = \int_{-\infty}^\infty dy\, E(y)\VEV{y\left|{1\over 1-z \bm\rho}\right|x}\,,  
\end{align}
where the reference state is defined as $E(x)=\sqrt{2}\, e^{x\over 2k}$. 
This function has a number of remarkable properties which we summarize below \cite{Tracy:1995ax,Boldis:2025yll}. 

It follows from the properties of the kernel \eqref{rho-fun} that the function $\psi(x|z)$ obeys the periodicity condition 
\begin{align}\label{period}
\psi(x+2\pi im k|z)=(-1)^m \psi(x|z)\,, \qqqquad m \in \mathbb{Z}\,.
\end{align}
In addition, the function $\psi(x|z)$ has an infinite set of logarithmic branch cuts in the complex $x-$plane that run parallel to the real axis and are located at $\Im x=(2m-1)\pi k$ for $m \in \mathbb Z$. The periodicity relation \re{period} allows us to restrict the analysis to the fundamental analyticity strip $-\pi k < \Im x< \pi k$.

Evaluating the discontinuity of the function \re{psi} across the branch cut and taking into account \re{period} leads to the  Baxter equation
\begin{align}\label{Bax}
\psi(x+i\pi k|z) + \psi(x-i\pi k|z) = {z \over 2\cosh(x/2)} \psi(x|z) \,,
\end{align}     
where $x$ is real and the functions on the right-hand side are evaluated at the boundary of the analyticity strip. 
The solution to \re{Bax} satisfies a Wronskian identity as well as the so-called shift relation \footnote{These relations are extensively discussed in Section 3 of \cite{Boldis:2025yll}. }, which allow us to express
$\psi(x+2\pi i|z)$ as a linear combination of the functions $\psi(x|z)$ and $\psi(x|-z)$.  

The rationale for introducing the function $\psi(x|z)$ is that the generating function $\mathcal W_n(\mu,k)$ in \re{det3} can be expressed in terms of $\psi(x|z)$ as follows~\footnote{The derivation of this relation can be found in Appendix A of \cite{Boldis:2025yll}.}  
\begin{align}\notag\label{W-psi}
\mathcal W_n(\mu,k) {}&= \frac{i z}{2\sin\left(\frac{2\pi n}{k}\right)} \frac{e^{i\pi n(n-1)/k}}{4\pi k} \int_{-\infty}^\infty 
\frac{dx \, e^{(n-1)x/k}}{2\cosh(x/2)} 
\\[1mm]
{}&\quad \times \Big[ e^{-i\pi n/k}\, \psi(x|z)\,\psi(x+2i\pi n|-z) - e^{i\pi n/k}\, \psi(x|-z)\,\psi(x+2i\pi n|z) \Big] \,.
\end{align}  
This relation is exact and holds for arbitrary values of $z=e^\mu$.  
 
In practice, evaluating $\mathcal W_n(\mu,k)$ from \eqref{W-psi} requires supplementing it with the solution to the Baxter equation \re{Bax}. Remarkably, as shown in \cite{Boldis:2025yll}, the relations satisfied by the functions $\psi(x|z)$ are sufficient to determine $\mathcal W_n(\mu,k)$ without the need for explicit expressions of $\psi(x|z)$. 
The resulting expression for $\mathcal{W}_n(\mu,k)$ depends on two functions, $\mathcal F(z)$ and $\Phi(z)$, defined as follows.

The first function, $\mathcal F(z)$, is obtained from the ratio of the generating functions $\Xi(\mu,k)$ under a shift of the chemical potential, $\mu \to \mu + i\pi$:
\begin{align}\label{F-def}
D(\mu) = e^{\mathcal F(z)} = \frac{\Xi(\mu,k)}{\Xi(\mu+i\pi,k)}, \qqqquad z = e^\mu.
\end{align}
Here and in what follows, the dependence of the functions on the Chern-Simons level $k$ is tacitly assumed.  
Since $\Xi(\mu,k)$ is invariant under $\mu \to \mu + 2i\pi$, the functions $D(\mu)$ and $\mathcal F(z)$ satisfy the relations
\begin{align}
D(\mu+i\pi) = \frac{1}{D(\mu)}, \qqqquad \mathcal F(-z) = -\mathcal F(z).
\end{align}
This property allows us to interpret $\mathcal F(z)$ as the parity-odd part of the grand potential in the Fermi--gas ensemble. Both $D(\mu)$ and $\mathcal F(z)$ are real-valued functions of their respective arguments.

The second function, $\Phi(z)$, is defined as the phase of the ratio of the Baxter functions $\psi(x|z)$ evaluated at $x = \pm i\pi$
\begin{align}\label{Phi-def}
\frac{\psi(i\pi|z)}{\psi(-i\pi|z)} = e^{i\Phi(z)}.
\end{align}
As we show in the following sections, the functions $\mathcal F(z)$ and $\Phi(z)$ are not independent and are related to each other through quantization conditions.   

For arbitrary values of the fugacity parameter $z = e^\mu$ and the winding number $n \geq 1$, the generating function \re{W-psi} takes the form \cite{Boldis:2025yll}
\begin{align}\label{W-cf}
\mathcal W_n(\mu,k)
= \frac{i^{\,n-1}}{\sin\!\left(\frac{2\pi n}{k}\right)}
\Big[
\mathcal A_n(z)
+ i\, z \partial_z \Phi(z)\, \mathcal B_n(z)
+ i\, z \partial_z \mathcal F(z)\, \mathcal C_n(z)
\Big].
\end{align}
The nontrivial dependence of $\mathcal W_n(\mu,k)$ on the winding number $n$ is encoded in the real functions $\mathcal A_n(z)$, $\mathcal B_n(z)$, and $\mathcal C_n(z)$. These functions, in turn, depend on the quantities $\mathcal F(z)$ and $\Phi(z)$ defined above.
For the first few values of $n$, they admit particularly compact expressions
\begin{itemize}

\item For $n=1$ 
\begin{align}\notag\label{A-n=1}
{}& \mathcal A_1(z)=\sinh\mathcal F(z)\sin\Phi(z)\,,
\\[2mm]\notag
{}& \mathcal B_1(z)=-\sinh\mathcal  F(z) \left[2 \cot (\ft{2 \pi }{k})\cos \Phi (z)+\sin \Phi (z)\right]\,,
\\[2mm]
{}& \mathcal C_1(z)=-\sinh\mathcal  F(z) \cos \Phi (z)-\cosh\mathcal  F(z) \left[2\cot (\ft{2\pi
   }{k}) \sin \Phi (z)-\cos \Phi (z)\right]\,.
\end{align}

\item 
For $n=2$ 
\begin{align}\notag\label{A-n=2}
 \mathcal A_2(z){}&=  (\cosh (2\mathcal  F(z))-1) \left(\cos^2 (\ft{2\pi }{k})-\cos (2 \Phi
   (z))\right)+\sinh (2\mathcal  F(z)) \sin ^2(\ft{2 \pi }{k}) \,,
\\[2mm]\notag
 \mathcal B_2(z){}&=(\cosh (2\mathcal  F(z))-1) \left[\cos (2 \Phi (z))-{2 \left(2 \cos (\ft{4 \pi }{k})+1\right)\over \sin (\frac{4 \pi }{k})}\sin (2 \Phi (z))-\ft{1}{2} (\cos(\ft{4
   \pi }{k})+3) \right]
\\{}&   \notag
+\sinh (2\mathcal  F(z))\cos ^2(\ft{2 \pi}{k}) \,,
\\[3mm]\notag
\mathcal C_2(z){}&=  - (\cosh (2\mathcal  F(z))-1) \left[\cos (\ft{4 \pi }{k}) \tan (\ft{2 \pi }{k})+ \sin (2 \Phi(z))\right]-
\cos (\ft{4 \pi }{k}) \tan (\ft{2 \pi }{k})
\\ {}&    
   +\sinh (2\mathcal  F(z)) \left[\sin (2 \Phi (z)) +{2 \left(2 \cos (\ft{4 \pi }{k})+1\right)\over \sin (\frac{4 \pi }{k})}\cos (2 \Phi
   (z))-\left(\cos (\ft{4 \pi }{k})+2\right) \cot (\ft{2 \pi}{k})\right].
\end{align}
\end{itemize}
For higher values of $n$, the coefficient functions follow the same structural pattern, but their explicit expressions rapidly grow in length \cite{Boldis:2025yll}.  

By definition \re{cal-W}, the function $\mathcal W_n(\mu,k)$ is the generating function of the unnormalized $1/6$ BPS Wilson loop. Using the relation \re{W-half}, we can define the corresponding generating function for  the $1/2$ BPS Wilson loop as
\begin{align}\label{W1/2-cf}
\mathcal W^{1/2}_n(\mu,k) = \mathcal W_n(\mu,k)-(-1)^n \overline{\mathcal W_n(\mu,k)}
= \frac{2\, i^{n-1}}{\sin(\frac{2\pi n}{k})}\mathcal A_n(z)\,.
\end{align}
The second relation follows from the fact that the last two terms in the bracket of \re{W-cf} are purely imaginary.

\subsection{Quantization conditions}

The generating functions \re{W-cf} and \re{W1/2-cf} depend on two auxiliary functions, $\mathcal F(z)$ and $\Phi(z)$, defined in \re{F-def} and \re{Phi-def}. Their determination proceeds in two steps. First, we establish the relation between these functions and, then, compute $\mathcal F(z)$.  Finally, in the next section we evaluate the Wilson loops.

As mentioned above, the solutions of the Baxter equation~\re{Bax} satisfy additional constraints in the form of Wronskian and shift relations. For positive integer values of~$k$, the requirement that these relations be mutually consistent leads to nontrivial constraints relating the functions $\mathcal F(z)$ and $\Phi(z)$. 

The resulting quantization conditions were derived in \cite{Boldis:2025yll}. 
For even $k$, they look as 
\begin{align}\label{QC0}
     (1,1)\, \mathbb T_{\rm e}  \lr{1 \atop 1}  = \frac12 z\,,
\end{align}
while for odd $k$ they take the form
\begin{align} \label{QC}
 \lr{e^{i\Phi/2},\ e^{-i\Phi/2}} \,(\mathbb{T}_{\rm o})^{\rm t} \sigma_3 \,\mathbb{T}_{\rm o} \,\lr{e^{i\Phi/2}\atop e^{-i\Phi/2}}= {i z^2 \over 2k\sinh \mathcal F(z)} \,,
\end{align}
where $\sigma_3$ is the Pauli matrix. Here $\mathbb T_{\rm e}$ and $\mathbb T_{\rm o}$ are $2\times 2$ matrices with the subscript referring to the parity of $k$.
For $k=1$ the matrix $\mathbb{T}_{\rm o}$ reduces to the identity matrix.
For $k\ge 2$, we have  
\begin{align}\notag
{}& \mathbb T_{\rm e} = T(i\pi(k-2)) \dots T(2i\pi)T(0)\,,
\\[2mm]
{}& \mathbb{T}_{\rm o} = T(i\pi(k-2)) \dots T(3i\pi)\,T(i\pi)\,,
\end{align}
where the matrix elements of $T(x)$ depend  nontrivially on the functions $\mathcal F(z)$ and $\Phi(z)$
\begin{align}\label{T-mat}
T(x) ={\displaystyle\sinh (\mathcal F(z))\over\displaystyle\sinh (\ft{x+i \pi }{k})} \left[
\begin{array}{cc} e^{i \Phi(z) } & -1\\ 1& -e^{-i \Phi(z) }  \\ \end{array} \right]+ e^{-\mathcal F(z)}\left[
\begin{array}{cc} 0 & e^{\frac{x+i \pi }{k}}\\ e^{-\frac{x+i \pi }{k}} & 0\\\end{array}\right].
\end{align}

Combining the above relations, we obtain from \re{QC0} and \re{QC} a functional relation between $\Phi(z)$ and $\mathcal F(z)$. For the first few values of the Chern--Simons level, $k=1,2,3,4$, the resulting
relations are, respectively,
\begin{align}\notag\label{QCs}
 z^2 ={}& -4\sinh \mathcal F \sin \widetilde\Phi\,,
\\[2mm]\notag
z\ ={}& 4 \sinh \mathcal F \cos \widetilde\Phi\,,
\\[2mm]\notag
z^2={}& -6 \sinh \mathcal F \sin \widetilde\Phi -12 \sinh \mathcal F\sin (3 \widetilde\Phi)+6 \sinh (3 \mathcal F)
   \sin \widetilde\Phi
\\[2mm]\notag
{}&    
   +4 \sinh (3 \mathcal F) \sin (3 \widetilde\Phi)-6 \sqrt{3} \cosh \mathcal F
   \cos \widetilde\Phi+6 \sqrt{3} \cosh (3 \mathcal F) \cos \widetilde\Phi\,,
\\[2mm] 
z\ ={}& 4 \cosh (2\mathcal F)
   \sin (2 \widetilde\Phi)+4 \sinh (2\mathcal F)-4 \sin (2\widetilde\Phi) ,
\end{align}
where the notation was introduced for $\widetilde \Phi(z) = \Phi(z)-\pi/ k$.  

The relations \re{QC0}, \re{QC} and \re{QCs} hold for arbitrary values of $z$ and positive integer $k$. 
Examining these relations, we find that $\mathcal F(z)$ has to grow at large $z$ as $\mathcal F(z)\sim 2\log z/k$. This suggests to look for a general solution for $D(\mu)=e^{\mathcal F(z)}$  at large $\mu$ of the form
\begin{align}\label{D-ans}
D(\mu)={e^{2\mu/k}\over 2\sin\Phi}\sum_{m\ge 0} q_m(\Phi) \, e^{-4m\mu/k} \,,
\end{align}
where the expansion coefficients $q_m(\Phi)$ depend on $k$. 

Substituting the ansatz \re{D-ans} into the quantization conditions \re{QC0}, \re{QC} and \re{QCs} and comparing the coefficients of powers of $e^{-4\mu/k}$ on both sides, we can determine the expansion coefficients for integer positive $k$. The first few coefficients are given by
\begin{align}\notag\label{gamma_lin}
q_0 =&1 \, , 
\\[2mm]\notag
q_1 =&2\delta _{1,k}+2\delta _{2,k}-2\cos 2\Phi\notag \, ,
\\[2mm]\notag
q_2 =&-\left(7 \delta _{1,k}+7 \delta _{2,k}+\delta _{4,k}-1\right)+\left(4\delta_{1,k}+4\delta_{2,k}+2+2\cos (\ft{4 \pi }{k})\right) \cos (2 \Phi )
\\[2mm]
&-2\sin \left(\ft{4 \pi }{k}\right) \sin (2 \Phi ) -2\cos (4 \Phi )\,.
\end{align}
Note that these expressions are not continuous in $k$ due to the presence of Kronecker delta functions $\delta_{s,k}$, with $s=1,2,4,\dots$. 

A careful analysis shows that the coefficients $q_m$ contain terms proportional to $\delta_{s,k}$ only when $\ell = 2m/s$ is an integer. An important observation in this case is that
$e^{-4m\mu/k} = e^{-2\ell \mu}$. This suggests to treat $e^{-4\mu/k}$ and $e^{-2\mu}$  as independent variables, and recast \re{D-ans} as a double series expansion in powers of these two parameters  
\begin{align}\label{D-gen}\notag
D(\mu) {}&= {e^{2\mu/k}\over 2\sin\Phi}\sum_{m,\ell\ge 0}\gamma_{m\ell}(\Phi) \, e^{-({4m\over k}+2\ell)\mu}
\\
{}&= {e^{2\mu/k}\over 2\sin\Phi}\lr{1+\gamma_{10}(\Phi) \, e^{- {4\mu\over k}}+\gamma_{01}(\Phi)\,e^{-2\mu}+\dots}\,,
\end{align}
where $\gamma_{00}=1$ and dots denote subleading terms. 

Comparing the two expansions \re{D-ans} and \re{D-gen}, we find that the two sets of the expansion coefficients, $q_m(\Phi)$ and $\gamma_{m\ell}(\Phi)$, are related to each other by linear relations. Examining these relations for various values of integer positive $k$, we obtained the following expressions for the coefficients $\gamma_{m\ell}(\Phi)$
\begin{align}\notag\label{g-WS}
\gamma_{10}&=-2\cos (2\Phi)\,,
\\[2mm]\notag
\gamma_{01}&=-\frac{4}{k}\cos\left(\frac{\pi k}{2}\right)\, ,  
\\\notag
\gamma_{11}&=-\frac{8}{k}\cos\left(\frac{\pi k}{2}\right)\cos (2\Phi)\,,
\\\notag
\gamma_{02}&= \frac{4}{k^2}-\frac{8}{k}+\left(\frac{4}{k^2}-\frac{10}{k} \right) \cos (\pi  k)\,, 
\\\notag
\gamma_{20}&=1+2 \cos \left(\frac{4 \pi }{k}+2 \Phi\right)+2\cos (2 \Phi )-2\cos (4 \Phi )\,,
\\\notag
\gamma_{12}&=-\left[\frac{8}{k^2}+\frac{16}{k}+\left(\frac{8}{k^2}+\frac{20}{k} \right) \cos (\pi  k)\right]\cos (2\Phi)\,,
\\
\gamma_{21}&=\frac{12}{k}\cos\left(\frac{\pi k}{2}\right)\left[1+2 \cos \left(\frac{4 \pi }{k}+2 \Phi \right) +2\cos (2 \Phi )-2\cos (4 \Phi )\right].
\end{align}
In distinction with \re{gamma_lin}, these relations do not contain Kronecker delta functions and define continuous functions of $k$. 

\subsection{Redefinition of the chemical potential}\label{sect:mueff}

Examining the coefficients \re{g-WS}, we observe that they are not independent and satisfy the relations
\begin{align}\label{rels}
\gamma_{11}  = -\gamma_{01}\gamma_{10}\,,\qqquad \gamma_{21}=-3\gamma_{01}\gamma_{20}\,,\qqquad \gamma_{12}=\gamma_{10}(\gamma_{01}^2-\gamma_{02})\,.
\end{align}
Moreover, the dependence on $\Phi$ cancels out in the ratio $\gamma_{m1}/\gamma_{m0}$ for $m=0,1,2$.
Analogous relations hold between the coefficients $\gamma_{m\ell}(\Phi)$ for higher values of $m$ and $\ell$.

As we will see in a moment, these relations stem from the following remarkable property of the expansion \re{D-gen}. It turns out that all terms in the double sum in \re{D-gen} with $\ell\ge 1$ can be absorbed into a redefinition the chemical potential $\mu\to \mu_{\rm eff}(\mu)$
\begin{align}\label{D-imp}
D(\mu)={e^{2\mu_{\rm eff}/k}\over 2\sin\Phi}\bigg[1+\sum_{m\ge 1}\gamma_{m0}(\Phi) \, e^{-{4m\over k} \mu_{\rm eff}}\bigg].
\end{align}
This relation elucidates the origin of the effective chemical potential in ABJM matrix model \re{Z}.

In order to reproduce the expansion \re{D-gen}, the effective chemical potential $\mu_{\rm eff}(\mu)$ must incorporate corrections organized in powers of $e^{-2\mu}$. This motivates the general form \re{eq:mueff} for $\mu_{\rm eff}(\mu)$. It is important to emphasize that we shall regard  \re{eq:mueff} as an ansatz, with the function $C(k)$ given by \re{CB} and the coefficients $a_\ell(k)$ treated as \emph{a priori} unknown. Our aim is to show that the relations \re{D-gen} and \re{g-WS} uniquely determine these coefficients.

Replacing $\mu_{\rm eff}$ in \re{D-gen} with the ansatz  \re{eq:mueff} and matching the coefficients of $e^{-({4m\over k}+2\ell)\mu}$ to those of the expansion \re{D-imp}, we can express $\gamma_{m\ell}(\Phi)$ with $\ell\ge 1$ in terms of $\gamma_{m0}(\Phi)$ and the unknown coefficients $a_\ell$. In this way, we find for $m\ge 0$ and $\ell \ge 1$
\begin{align}
\widehat \gamma_{m\ell} \equiv {\gamma_{m\ell}\over \gamma_{m0}}=\sum_{p_1,\dots,\,p_\ell\ge 0} {(-(2m-1) \pi^2 a_1)^{p_1} \dots (-(2m-1) \pi^2a_\ell)^{p_\ell}\over p_1! \dots p_\ell!}\,,
\end{align}
where the sum runs over nonnegative integers satisfying $p_1+2p_2+\dots + \ell p_\ell=\ell$. 
For lowest values of $\ell=1,2$ this relation looks as
\begin{align}\notag
{}& \widehat \gamma_{m1} =-(2m-1) \pi^2 a_1\,,
\\[2mm] 
{}& \widehat \gamma_{m2} =\frac12 \lr{(2m-1) \pi^2 a_1}^2-(2m-1) \pi^2 a_2\,.
\end{align}
We can invert these relations to express the $a-$coefficients in terms of $\widehat\gamma$'s
\begin{align}\notag\label{a-gamma} 
 {}& a_1=-{1\over (2m-1) \pi^2}\widehat \gamma_{m1}\,, 
 \\
 {}& a_2=-{1\over (2m-1) \pi^2}\lr{\widehat \gamma_{m2}-\frac12 \widehat \gamma_{m1}^2}\,.
\end{align}
Notice that these relations holds for arbitrary $m\ge 0$. For the left-hand side to remain independent of $m$, the coefficients  $\widehat \gamma_{m\ell}=\gamma_{m\ell}/\gamma_{m0}$ must satisfy a set of nontrivial constraints. For the lowest values of $m$, these constraints reduce precisely to the relations \re{rels}. The cancellation of the $\Phi-$dependence in $\widehat \gamma_{m\ell}$ guarantees that the coefficients 
$a_\ell$  are independent of $\Phi$.

In addition, the relations \re{a-gamma} can be used to determine the coefficients $a_\ell$. Substituting $m=0$ and replacing $\gamma$'s with their expressions \re{g-WS} we arrive at
\begin{align}\notag\label{aa}
    &a_{1}(k)=-\frac{4 }{\pi ^2 k}\cos \left(\frac{\pi  k}{2}\right),\\
    &a_2(k)=-\frac{2}{\pi ^2 k}(4+5 \cos (\pi  k))\, .
\end{align}
We immediately recognize that these results agree with the conjectured expressions for $a_1$ and $a_2$ in \re{AB}. 
The same procedure can be straightforwardly extended to the computation of coefficients $a_\ell$ for higher $\ell$. Indeed, we have calculated $a_\ell$ up to $\ell = 5$ 
and verified that our results coincide with those reported in \cite{Hatsuda:2013gj}. This completes the evaluation of the effective chemical potential \re{eq:mueff}.

The relation \re{D-imp} involves the set of coefficient functions $\gamma_{m0}(\Phi)$. For $m=1,2$ these coefficients are given by \re{g-WS}. For higher values of $m$, they take the form
\begin{align}\label{g0}
    \gamma_{m,0}(\Phi)=\sum_{r=0}^m\sum_{s=0}^{\lfloor m^2/4 \rfloor } \alpha_{r,s}\cos\left(2r\Phi\right)\cos\left(\frac{4s\pi}{k}\right)+  \beta_{r,s}\sin\left(2r\Phi\right)\sin\left(\frac{4s\pi}{k}\right),
\end{align}
where $\alpha_{r,s}$ and $\beta_{r,s}$ are integers. Explicit expressions for $\gamma_{m,0}(\Phi)$ up to $m=9$ are provided in the attached ancillary file. 

Combining the relations \eqref{D-imp} and \eqref{g0} allows us to express the function $\Phi(z)$ in terms of $D(\mu)$. In the next step, we proceed to evaluate the function $D(\mu)$. 
  
\subsection{Nonperturbative contribution} 
 
By definition \re{F-def}, the function $D(\mu)$ is given by the ratio of grand canonical partition functions $\Xi(\mu,k)$ evaluated at shifted arguments. Replacing $\Xi(\mu,k)$ with its representation \re{Xi-conj}, we can rewrite $D(\mu)$ in terms of the grand potential as
\begin{align}\label{Dfra}
D(\mu) = \frac{\displaystyle \sum_{m=-\infty}^{\infty} 
      e^{J(\mu + 2\pi i m,k) - J(\mu,k)}}%
     {\displaystyle \sum_{m=-\infty}^{\infty} 
      e^{J(\mu + \pi i (2m-1),k) - J(\mu,k)}} \,.
\end{align}
Here $J(\mu,k)$ was subtracted from the exponents of both numerator and denominator for convenience, as it cancels in the ratio.

As discussed in Section \ref{sec2}, the grand potential $J(\mu,k)$ is given by the general expression \re{J-grand}. Its perturbative part \re{J-pt} follows from the semiclassical expansion of the Fermi gas, while the nonperturbative contribution \re{J-eff} was inferred from a network of conjectures, which we now aim to derive. By relaxing these conjectures, we seek a general expression for $J(\mu,k)$ of the form
\begin{align}\notag\label{J-gen}
J(\mu,k) {}& = \sum_{m,\ell\ge0} \widetilde f_{m,\ell}(\mu_{\rm eff})\, e^{-\big({4m\over k}+2\ell\big)\mu_{\rm eff}}
\\
{}&  = \frac13 C(k) \mu_{\rm eff}^3 + B(k)\mu_{\rm eff} + A(k)+ \sum_{m+\ell\ge 1} \widetilde f_{m,\ell}(\mu_{\rm eff})\,Q_{\rm w}^m \,Q_{\rm m}^\ell\,,
\end{align}
where $\widetilde f_{0,0}(\mu)$ corresponds to the perturbative part \re{CB} and the effective chemical potential $\mu_{\rm eff}$ is given by \re{eq:mueff}. 
In the second relation, the notation was introduced for  
\begin{align}\label{Q-def}
Q_{\rm w}=e^{-4\mu_{\rm eff}/k}\,,\qqqquad Q_{\rm m}=e^{-2\mu_{\rm eff}}\,,
\end{align}
which encode the contributions of world-sheet and membrane instantons in the dual description (motivating the choice of subscripts).  In what follows, we treat \re{J-gen} as an ansatz and regard the coefficient functions $\widetilde f_{m,\ell}(\mu_{\rm eff})$ as unknowns to be determined.

To prove the conjectures formulated in Section \ref{sec2}, it remains to show that the coefficients in \re{J-gen} satisfy 
\begin{align}\notag
{}&\widetilde f_{0,\ell}(\mu_{\rm eff})=\mu_{\rm eff}\,\tilde b_\ell(k)+\tilde c_\ell(k)\,,
\\[2mm]\notag
{}&\widetilde f_{m,0}(\mu_{\rm eff})=d_m\,,
\\[2mm]\label{f-conj}
{}& \widetilde f_{m,\ell}(\mu_{\rm eff})=0 \,, 
\end{align}
for $m\,, \ell\geq 1$. The coefficients in the first two relations are defined in \re{JM2} and \re{fm0}, respectively, while the last relation reflects the absence of the bound state contribution in \re{J-eff}. 
   
Replacing $J(\mu,k)$ in \re{Dfra} with the ansatz \re{J-gen}, we can express the function $D(\mu)$ in terms of the coefficient functions $\widetilde f_{m,\ell}(\mu_{\rm eff})$. It follows from \re{eq:mueff} that the shift of the chemical potential $\mu\to\mu+i\pi$ corresponds to $\mu_{\rm eff}\to\mu_{\rm eff}+i\pi$. We use this property to find from \re{J-gen}
\begin{align}\label{osc}
e^{J(\mu+im\pi,k)-J(\mu,k)} = Q_{\rm w}^{m^2/2} \exp\lr{im\lr{{\pi\over 2}- \varphi(\mu_{\rm eff})}-{2i\pi\over 3k}m(m^2-1) +O(Q_{\rm w},Q_{\rm m})}\,,
\end{align}
where $m$ is integer, the parameter $Q_{\rm w}$ is defined in \re{Q-def} and the notation was introduced for the phase
\begin{align} \label{varphi}
 \varphi(\mu_{\rm eff}) =- \frac{2}{\pi k}\mu^2_{\mathrm{eff}}+\frac{\pi}{3k}+\frac{\pi}{2}-\frac{\pi k}{24}\,.
\end{align} 
Note that this phase grows quadratically at large $\mu_{\mathrm{eff}}$ and, as a consequence, the exponential factor in \re{osc} is a rapidly oscillating function of $\mu_{\mathrm{eff}}$. This property plays an important role in what follows. 
 
Due to the factor $Q_{\rm w}^{m^2/2}$ on the right-hand side of \re{osc}, the contributions from terms with large $m$ in both the numerator and denominator of \re{Dfra} are subleading at large $\mu$. In particular, keeping in \re{Dfra} the $m=0$ term in the numerator and the $m=0$ and $m=1$ terms in the denominator we get
\begin{align}\notag\label{ab}
D(\mu) {}&= \frac{1}{%
        e^{J(\mu+i\pi,k)-J(\mu,k)} + e^{J(\mu-i\pi,k)-J(\mu,k)}} 
        + \dots \,.
        \\
        {}&= {e^{2\mu_{\rm eff}/k}\over 2\sin\varphi(\mu_{\rm eff})} \left[ 1+ O(Q_{\rm w})+O(Q_{\rm m})\right],
\end{align}
where in the second relation we took into account \re{osc}. 

The relation \re{ab} is remarkably similar to \re{D-imp}. By comparing the two relations, we find that $\Phi(\mu)$ coincides with $\varphi(\mu_{\rm eff})$ up to subleading corrections. This allows us to write a general expression for the functions $D$ and $\Phi$ in the form
\begin{align}\notag\label{Phi-exp}
{}& D(\mu) ={e^{2\mu_{\rm eff}/k}\over 2\sin\varphi(\mu_{\rm eff})}  \bigg[1+\sum_{m+\ell\ge 1} D_{m\ell}(\mu_{\rm eff}) Q_{\rm w}^m Q_{\rm m}^\ell\bigg],
\\[1.2mm]
{}& \Phi(\mu) = \varphi(\mu_{\rm eff})+\sum_{m+\ell\ge 1} \phi_{m\ell}(\mu_{\rm eff}) Q_{\rm w}^m Q_{\rm m}^\ell\,,
\end{align}
where the sum runs over nonnegative integers except $m=\ell=0$. \footnote{As we show below (see \re{Phi-imp}), the coefficients $\phi_{m\ell}(\mu_{\rm eff})$ are not independent. They can be expressed in terms of $\phi_{m0}(\mu_{\rm eff})$ and the membrane coefficients $\tilde b_\ell$.}

It was shown in \cite{Boldis:2025yll} that, upon substituting the expressions \re{Phi-exp} and \re{varphi} into the generating function~\re{W-cf} and neglecting nonperturbative corrections, one precisely recovers the expected result~\re{W-pert} for the perturbative contribution to the Wilson loop.

To compute the nonperturbative contribution to the generating function \re{W-cf}, it is necessary to determine the subleading corrections in \re{Phi-exp}. The coefficients $D_{m\ell}(\mu_{\rm eff})$ can be obtained by retaining additional terms in \re{Dfra} and expanding the resulting expression for $D(\mu)$ in powers of the parameters defined in \re{Q-def}. These coefficients  depend explicitly on the functions $\widetilde f_{m,\ell}(\mu_{\rm eff})$, which parametrize the nonperturbative corrections to the grand potential \re{J-gen}.
For instance, the first two coefficients are given by
\begin{align} \notag\label{D01}
{}& D_{01} = \pi \cot(\varphi ) {\widetilde f}_{01}^{\,\prime}(\mu_{\rm eff})
\\[2mm]
{}& D_{10}=  \left(1- {\sin \left(\frac{4 \pi }{k}+\varphi\right)\over \sin(\varphi)}\right){\widetilde f}_{10}(\mu_{\rm eff})+\pi   {\cos \left(\frac{4 \pi }{k}+\varphi  \right) \over \sin (\varphi)}{\widetilde f}_{10}^{\,\prime}(\mu_{\rm eff}) \,,
\end{align}
where $\varphi=\varphi(\mu_{\rm eff})$ and prime denote a derivative with respect to $\mu_{\rm eff}$. Here, for the sake of simplicity, we assumed that $f_{m\ell}(\mu_{\rm eff})$ are linear functions of $\mu_{\rm eff}$. \footnote{As we show below, this assumption is justified a posteriori.} We observe that the functions $D_{01}$ and $D_{10}$ depend on the phase \re{varphi} and rapidly oscillate at large $\mu_{\rm eff}$. The same holds for the higher-order coefficients $D_{m\ell}$ in \re{Phi-exp}.

Similarly, combining the relations \re{D-imp} and \re{g-WS} and substituting the ansatz \re{Phi-exp} for the phase $\Phi$, we can express the expansion coefficients $\phi_{m\ell}$ in terms of the $D$’s. Using \re{D01}, these coefficients can then be written in terms of $\widetilde f_{m\ell}(\mu_{\rm eff})$. For instance,
\begin{align}\notag\label{phi's}
{}& \phi _{01}=-\pi  {\widetilde f}_{01}^{\,\prime}(\mu_{\rm eff})\,,
\\[2mm]
{}& \phi _{10}=-\tan(\varphi) \left[2\cos (2\varphi)+\left(1- {\sin \left(\frac{4 \pi }{k}+\varphi\right)\over \sin(\varphi )}\right){\widetilde f}_{10}(\mu_{\rm eff})+\pi   {\cos \left(\frac{4 \pi }{k}+\varphi \right) \over \sin (\varphi )}{\widetilde f}_{10}^{\,\prime}(\mu_{\rm eff})\right].
\end{align}
In close analogy with \re{D01}, the coefficients $\phi_{m\ell}$ depend on the phase \re{varphi} and exhibit rapid oscillations at large $\mu_{\rm eff}$.

Finally, inserting the expansions \re{Phi-exp} for the functions $\mathcal F=\log D(\mu)$ and $\Phi(\mu)$ into \re{W-cf}, and using \re{D01} together with \re{phi's}, we can compute the nonperturbative corrections to the generating function $\mathcal W_n(\mu,k)$ in terms of $\widetilde f_{m\ell}(\mu_{\rm eff})$ and their derivatives. These functions parameterize the nonperturbative contributions to the grand potential~\re{J-gen} and remain undetermined at this stage. In the next section, we determine them explicitly.
  
\section{Bootstrap approach}
\label{sec:4}

In the previous sections, we computed the generating function
$\mathcal W_n(\mu,k)$ in the large-$\mu$ regime. 
Substituting this result into~\re{W-GC} enables us to invert the grand-canonical ensemble
average and determine the function $\mathsf W^{1/6}_n(\mu,k)$.
The expectation value of the $1/6$~BPS Wilson loop then follows by applying the integral
transform~\re{W-W}.

\subsection{Consistency condition}
 
Our subsequent analysis relies on the following observation concerning
the structure of the grand-canonical averaging~\re{vev-GC}. For an
arbitrary function $\mathsf U(\mu)$ of the chemical potential, its
grand-canonical average is defined as 
\begin{align}\label{vev-GC2}
\mathcal U(\mu)= \vev{U(\mu)}_{_{\rm GC}}
\equiv \frac{\displaystyle \sum_{m=-\infty}^{\infty}
\mathsf U(\mu+2\pi i m)\,
e^{J(\mu+2\pi i m,k)-J(\mu,k)}}%
{\displaystyle \sum_{m=-\infty}^{\infty}
e^{J(\mu+2 \pi i m,k)-J(\mu,k)}} \, .
\end{align}
Both the numerator and the denominator involve the exponential factors
\re{osc}, which introduce an explicit dependence of $\mathcal U(\mu)$ on the phase
$\varphi(\mu_{\rm eff})$ defined in~\re{varphi}.

As a consequence, even if $\mathsf U(\mu)$ is a slowly varying function of $\mu$, its grand-canonical average $\mathcal U(\mu)$ generically inherits a dependence on $\varphi(\mu_{\rm eff})$ and therefore becomes a rapidly oscillating function of $\mu$. Conversely, given a rapidly oscillating function $\mathcal U(\mu)$, there is no a priori guarantee that the corresponding function $\mathsf U(\mu)$ is slowly varying. Requiring this property to hold imposes a nontrivial consistency condition on the functions $\mathcal U(\mu)$ and $J(\mu,k)$. It is this condition that we shall exploit in the following.

More precisely, we require that, for arbitrary winding number, the function $\mathsf W^{1/6}_n(\mu,k)$ satisfying the relation \re{vev-GC} has the general form \re{WW-gen} -- \re{W-np} with the coefficient functions $\mathsf w_{m,\ell}(\mu)$ being slowly varying function of $\mu_{\rm eff}$ and independent of the phase $\varphi(\mu_{\rm eff})$.
 
As a warm-up example, let us consider the generating function of the
$1/2$~BPS Wilson loop~\re{W1/2-cf} with the winding number $n=1$,
\begin{align} \label{calW-1/2large}
\mathcal W^{1/2}_1(\mu,k) &= \frac{2}{\sin(\frac{2\pi}{k})}\, \mathcal A_1(z) = {\sin \Phi(\mu)\over \sin(\frac{2\pi}{k})}\lr{D(\mu)-{1\over D(\mu)}}\,,
\end{align}
where in the second relation we used \re{A-n=1}. Replacing the function $D(\mu)$ with its expression \re{D-imp}, we can expand $\mathcal W^{1/2}_1(\mu,k)$ in powers of $Q_{\rm w}$
\begin{align} \label{A1-exp}\notag
\mathcal W^{1/2}_1(\mu,k) &=\frac{e^{\frac{2\mu_{\mathrm{eff}}}{k}}}{2\sin(\frac{2\pi}{k})}\bigg[1-2 Q_{\rm w}+\lr{3-4 \sin \left(\ft{2 \pi }{k}\right) \sin\left(\ft{2\pi }{k}+2\Phi\right)}Q_{\rm w}^2 -\big(6+4 \cos \left(\ft{4\pi}{k}\right) 
\\
{}&-8 \sin \left(\ft{2 \pi }{k}\right) \left(2 \sin \left(\ft{2\pi }{k}+2\Phi
   \right)+\sin \left(\ft{6 \pi }{k}+2 \Phi \right)-\sin \left(\ft{2 \pi
   }{k}+4 \Phi \right)\right)\!\big)Q_{\rm w}^3+ O(Q_{\rm w}^4)\bigg].
\end{align}
It is convenient to regard $\mathcal W^{1/2}_1(\mu,k)$ as a function of the effective chemical potential~\re{eq:mueff} and of the parameters $Q_{\rm w}$ and $Q_{\rm m}$ defined in~\re{Q-def}. The dependence on $Q_{\rm m}$ enters \re{A1-exp} entirely through the phase $\Phi$, whose large-$\mu$ expansion is given in~\re{Phi-exp}.

Substituting the expansion \re{Phi-exp} into \re{A1-exp}, we obtain the following expansion
\begin{align}\label{calW-1/2gen}
\mathcal W^{1/2}_1(\mu,k) &=\frac{e^{\frac{2\mu_{\mathrm{eff}}}{k}}}{2\sin\left(\frac{2\pi}{k}\right)}\Big[1+\sum_{m\ge 1,\, \ell\ge 0}
 w_{m,\ell}(\mu_{\mathrm{eff}})\,Q_{\rm w}^mQ_{\rm m}^\ell\Big].
\end{align}
Note that the double sum does not contain the term with $m=0$. 
The first few expansion coefficients are given by
\begin{align}\notag\label{w's} 
{}& w_{10}=-2\,,\qquad  w_{11}=w_{12}=0\,,
\\[2mm]\notag
{}& w_{20}=-4 \sin \left(\ft{2 \pi }{k}\right) \sin\left(\ft{2\pi }{k}+2\varphi\right)+3\,,
\\[2mm] 
{}& w_{21}=-8 \sin \left(\ft{2 \pi }{k}\right) \cos\left(\ft{2\pi }{k}+2\varphi\right)\phi_{01}\,.
\end{align}
They depend on the phase \re{varphi} through the functions $\phi_{01}$ defined in \re{phi's}. In agreement with our expectations, the coefficients \re{w's} are rapidly oscillating at large $\mu_{\rm eff}$.

Following \re{vev-GC}, we identify the generating function \re{calW-1/2gen} as the grand canonical average
\begin{align}\label{vev-GC1}
 \mathcal W^{1/2}_1(\mu,k) = \vev{\mathsf W^{1/2}_1(\mu,k)}_{_{\rm GC}}\,.
\end{align}
The function $\mathsf W^{1/2}_1(\mu,k)$ has the form similar to \re{calW-1/2gen}
\begin{align}\label{sW-1/2gen}
\mathsf W^{1/2}_1(\mu,k) &=\frac{e^{\frac{2\mu_{\mathrm{eff}}}{k}}}{2\sin\left(\frac{2\pi}{k}\right)}\Big[1+\sum_{m\ge 1,\, \ell\ge 0}
 \mathsf w_{m,\ell}(\mu_{\mathrm{eff}})\,Q_{\rm w}^mQ_{\rm m}^\ell\Big],
\end{align}
where the expansion coefficients are related to those in \re{calW-1/2gen} by linear relations.
To find these relations we substitute the expansions \re{calW-1/2gen} and \re{sW-1/2gen} into \re{vev-GC1} and compare the coefficients of powers of $Q_{\rm w}$ and $Q_{\rm m}$ on both sides. 

In this way, we find that the coefficients of terms linear in $Q_{\rm w}$ and $Q_{\rm m}$ coincide, $w_{10}=\mathsf w_{10}$ and $w_{11}=\mathsf w_{11}$.
The remaining coefficients are related to each other as
\begin{align} \notag\label{sw's}
w_{20}{}& =\mathsf w_{20}-4\sin\lr{\ft{2\pi}{k}}\sin\lr{\ft{2\pi}{k}+2\varphi}\,,
\\[2mm] 
w_{21}{}&=\mathsf w_{21} +8\pi\sin\lr{2\pi\over k}\cos\lr{{2\pi\over k}+2\varphi}\widetilde f_{01}^{\,\prime}
.
\end{align}
Comparing these relations with \re{w's} and taking into account \re{phi's}, we determine the expansion coefficients
\begin{align}\label{res1}
\mathsf w_{11}=\mathsf w_{21}=0\,,\qqquad
\mathsf w_{10}=-2\,,\qqquad
\mathsf w_{20}=3\,.
\end{align}

The above analysis can be extended to derive two sets of relations for the coefficients
$w_{m,\ell}$, generalizing \re{w's} and \re{sw's}. Equating them we can obtain the system of linear equations for the  corresponding coefficients $\mathsf w_{m,\ell}$. In particular, the coefficient $\mathsf w_{30}$ satisfies the equation
\begin{align}\label{w30}
\mathsf w_{30}=- 10-\Omega_1(\varphi)\, \widetilde f_{10}^{\,\prime}  - \Omega_2(\varphi)\lr{\sin^2\!\lr{\ft{2\pi}{k}}\,\widetilde f_{10}-1}\,.
\end{align}
It involves two different functions $\Omega_i(\varphi)$  of the phase \re{varphi}. The explicit form of these functions is not important for our purposes. It is important, however, that these functions are independent of each other and rapidly oscillate at large $\mu_{\rm eff}$.

As explained above, the expansion coefficients in~\re{sW-1/2gen} are required to be slowly varying functions of $\mu_{\rm eff}$ and, consequently, must be independent of the phase $\varphi(\mu_{\rm eff})$. This requirement implies that the coefficients multiplying the functions $\Omega_i(\varphi)$ in \re{w30} have to vanish. One therefore immediately obtains
\begin{align}\label{res2}
\mathsf w_{30} = -10\,, \qqqquad
\widetilde f_{10} = \frac{1}{\sin^2\!\left(\frac{2\pi}{k}\right)} \, .
\end{align}
We recall that, in deriving the relation~\re{D01}, we assumed that $f_{01}(\mu_{\rm eff})$ and $f_{10}(\mu_{\rm eff})$ are linear functions of $\mu_{\rm eff}$. Relaxing this assumption would lead to additional contributions to \re{sw's} and~\re{w30}, involving higher derivatives of these functions multiplied by nontrivial functions of $\varphi$. Requiring the $\mathsf w-$coefficients to remain independent of $\varphi$ then forces all higher derivatives to vanish, thereby justifying the assumed linearity.

We would like to emphasize that the above procedure allows us to determine not only the Wilson loop coefficients \re{res1} and \re{res2}, but also the coefficient $\widetilde f_{10}$ parametrizing the nonperturbative correction to the grand potential~\re{J-gen}. We verify that the obtained result for $\widetilde f_{10}$ agrees with its conjectured value $\widetilde f_{10}=d_1(k)$ (see \re{f1f2f3} and \re{f-conj}).

The above analysis can be systematically extended to determine the remaining coefficients $\mathsf w_{m,\ell}$.  
For $m\ge 3$ and $\ell\ge 0$, these coefficients satisfy equation of the form  
\begin{align}\label{sys-eqs}
\mathsf w_{m\ell} + \Omega_1(\varphi) \widetilde f_{m-2,\ell}' +  \Omega_2 (\varphi) \widetilde f_{m-2,\ell} + \Omega_3 (\varphi) = 0\,,
\end{align}
where the coefficient functions $\Omega_i(\varphi)$ are, in general, different from those in \re{w30}. They depend on 
the coefficients $\mathsf w_{m'\ell'}$ and $\widetilde f_{m'-2,\ell'}$ with smaller values of indices. 

For $\ell=0$, the functions $\Omega_2(\varphi)$ and $\Omega_3(\varphi)$ are affinely dependent, $\Omega_3(\varphi)+d_{m-2}\,\Omega_2(\varphi) +\beta_{m}=0$, and are independent of the function $\Omega_1(\varphi)$. The corresponding solution to \re{sys-eqs} is therefore $\mathsf w_{m0}=\beta_m$ and $\widetilde f_{m-2,0}=d_{m-2}$. 
For $\ell\ge 1$, the function $\Omega_3(\varphi)$ vanishes and the equation \re{sys-eqs} implies
\begin{align}\label{zeros}
\mathsf w_{m\ell} =\widetilde f_{m-2,\ell}=0\,, 
\end{align}
for $m\ge 3$ and $\ell\ge 1$. 
 
\subsection{$1/2$ BPS Wilson loop} 

 Carrying out the procedure described in the previous subsection, we arrive at the following result for the generating function~\re{sW-1/2gen},
\begin{align}\label{W-n=1}
\mathsf W^{1/2}_1(\mu,k)
=\frac{e^{\frac{2\mu_{\mathrm{eff}}}{k}}}{2\sin\!\left(\ft{2\pi}{k}\right)}
\Big[
&1-2 \,Q_{\rm w}+3 \,Q_{\rm w}^2-10 \,Q_{\rm w}^3
+\left(33+16 \cos\!\left(\ft{4\pi}{k}\right)\right) Q_{\rm w}^4
\nonumber\\
&-\left(132+112 \cos\!\left(\ft{4\pi}{k}\right)
+44 \cos\!\left(\ft{8\pi}{k}\right)\right) Q_{\rm w}^5
+\mathcal O(Q_{\rm w}^6)
\Big].
\end{align}
The explicit expressions for the subleading terms up to $\mathcal O(Q_{\rm w}^9)$ are provided in the attached ancillary file. It is straightforward to verify that the relation \re{W-n=1} agrees with the analogous expression conjectured in \cite{Hatsuda:2013yua} (see (6.17) there).~\footnote{In the ’t Hooft limit, upon neglecting the membrane instanton corrections, the relation \re{W-n=1} agrees with the result of \cite{Grassi:2013qva} (see (4.13) therein).}

We observe that the expansion in~\re{W-n=1} involves powers of $Q_{\rm w}$ only. Consequently, all coefficients multiplying powers of $Q_{\rm m}$
in~\re{sW-1/2gen} vanish, see \re{zeros} and footnote ${}^{\ref{foot}}$. We show below that this property holds for the $1/2$~BPS Wilson loop $\mathsf W^{1/2}_n(\mu,k)$ for arbitrary winding number and to all orders in $Q_{\rm w}$. This result is consistent with the conjecture~\re{W-1/2-conj} put forward in~\cite{Hatsuda:2013yua}.

In addition to the result~\re{W-n=1}, it follows from \re{zeros} that the only nonvanishing $\widetilde f-$coefficients are $\widetilde f_{m,0}$ and $\widetilde f_{0,\ell}$ with $m,\ell\ge 1$. We explicitly computed the world-sheet coefficients $\widetilde f_{m,0}$ for $m\leq 7$ and found perfect agreement with the conjectured relations \re{f-conj}. The membrane coefficients $\widetilde f_{0,\ell}$ have to be linear functions of the effective chemical potential, $\widetilde f_{0,\ell}=\mu_{\rm eff}\,\tilde b_\ell(k)+\tilde c_\ell(k)$. The expansion coefficients $\tilde b_\ell(k)$ and $\tilde c_\ell(k)$
decouple from the equations determining the functions $\mathsf w_{m,\ell}$ and, consequently, remain undetermined within the present analysis. 
 
In close analogy with the effective chemical potential \re{eq:mueff}, the properties of $\mathsf W^{1/2}_n(\mu,k)$ can be elucidated by refining the phase \re{varphi} to incorporate non-perturbative corrections in $Q_{\rm m}$.
Taking into account the relations \re{J-gen} and \re{f-conj}, we can repeat the
calculation of~\re{osc}, now including nonperturbative corrections to
the grand potential in $Q_{\rm m}$,
\begin{align}\label{osc1} 
e^{J(\mu+im\pi,k)-J(\mu,k)}=Q_{\rm w}^{m^2/2} \exp\lr{im\lr{{\pi\over 2}- \varphi+ \pi \widetilde J_b(\mu_{\rm eff})}-{2i\pi\over 3k}m(m^2-1) +O(Q_{\rm w})}\,,
\end{align}
where the function $\widetilde J_b(\mu_{\rm eff})$ was defined in \re{JM2}. Note that the dependence on the coefficients $\widetilde c_\ell(k)$ cancels out in \re{osc1}.
   
We observe that the dependence of the exponent in~\re{osc1} on $Q_{\rm m}$ resides in $\widetilde J_b(\mu_{\rm eff})$ and can therefore be absorbed into a redefinition of the phase,
\begin{align}\label{tildephi}
\widetilde\varphi
=\varphi(\mu_{\rm eff})-\pi\,\widetilde J_b(\mu_{\rm eff}) \, .
\end{align}
Replacing $\varphi$ by $\widetilde\varphi$ thus removes all contributions proportional to $Q_{\rm m}$ from the generating function $\mathcal W^{1/2}_n(\mu,k)$. Importantly, this redefinition leaves the coefficients $\mathsf w_{m,\ell}$ and $\widetilde f_{m,\ell}$ unchanged, since by construction they are independent of the phase $\varphi$. This explains why the large-$\mu$ expansion of the $1/2$~BPS Wilson loop $\mathsf W^{1/2}_n(\mu,k)$ contains no terms proportional to the membrane instanton parameter $Q_{\rm m}$.

Exploiting this property significantly simplifies the computation of $\mathsf W^{1/2}_n(\mu,k)$. For an arbitrary winding number $n\ge 1$ this function takes the form
\begin{align}\label{w-r-b}
    \mathsf W^{1/2}_n(\mu,k)=\frac{i^{n-1} e^{\frac{2n\mu_{\mathrm{eff}}}{k}}}{2\sin\left(\frac{2\pi n}{k}\right)}\Big(1+\sum_{m\ge 1} 
    \mathsf w_{m} Q_{\rm w}^m\Big)\,,
\end{align}
where the expansion coefficients $\mathsf w_m$ depend on $n$ and $k$, but are independent of the chemical potential. Imposing the condition $\vev{\mathsf W^{1/2}_n(\mu,k)}_{_{\rm GC}}=  \mathcal W^{1/2}_n(\mu,k)$ and using \re{W1/2-cf}, we find that the coefficients $\mathsf w_m$ obey
\begin{align}\label{eq1/2}
\VEV{ e^{\frac{2n\mu_{\mathrm{eff}}}{k}}\sum_{m\ge 1} 
    \mathsf w_m Q_{\rm w}^m}_{^{\rm GC}} = 4\mathcal A_n(z)\,,
\end{align}
where the grand canonical average is defined in \re{vev-GC2} and the coefficient functions $\mathcal A_n(z)$ are introduced in \re{W-cf}.  

Using the explicit expressions for the functions $\mathcal A_n(z)$ with $n\le 6$ obtained in~\cite{Boldis:2025yll}, we solved~\re{eq1/2} and computed the corresponding $1/2$~BPS Wilson loops $\mathsf W^{1/2}_n(\mu,k)$.  
We find that the leading nonperturbative correction takes the form
\begin{align}\label{lead}
\mathsf W_n^{1/2}(\mu,k)
= \frac{i^{\,n-1} e^{\frac{2n\mu_{\rm eff}}{k}}}{2\sin\!\left(\frac{2\pi n}{k}\right)}
\Biggl[
1 - \Bigl(\delta_{n,1} + \frac{\sin^2\!\left(\frac{2\pi n}{k}\right)}{\sin^2\!\left(\frac{2\pi}{k}\right)}\Bigr) Q_{\rm w}
+ O(Q_{\rm w}^2)
\Biggr].
\end{align}
We verified that, in agreement with the conjectured relations~\eqref{WS-conj} and~\eqref{W-1/2-conj}, this Wilson loop admits the equivalent representation
\begin{align}
\label{eq:1/2Winst}
\mathsf W_n^{1/2}(\mu,k)
= \frac{i^{n-1}e^{\frac{2n\mu_{\rm eff}}{k}}}{2\sin\!\frac{2\pi n}{k}} f((-Q_{\rm w})^n)
\sum_{g,d\ge 0} \sum_{\ell m=n} N_{\vec e_m,d}^g
\left(2\sin\frac{2\pi \ell}{k}\right)^{2g-2}
\left(2\sin\frac{2\pi n}{k}\right)^2
(-Q_{\rm w})^{d\ell}\,,
\end{align}
and computed the corresponding integers $N_{\vec e_m,d}^g$ up to $g=14$ and $d=9$. Their values, together with the subleading corrections to \re{lead}, are provided in the ancillary \texttt{Mathematica} file.

We emphasize that, as in the $n=1$ case, the equation~\re{eq1/2} also determines the coefficients $\widetilde f_{m,\ell}$ appearing in the grand potential~\eqref{J-gen}. As a nontrivial consistency check, we verified that these coefficients satisfy the relations \re{f-conj} and are independent of the winding number $n$, as required.

\subsection{$1/6$ BPS Wilson loop}
 
As explained in Section~\ref{sec2}, the nonperturbative corrections to $\mathsf W_n^{1/6}(\mu,k)$ exhibit a significantly more intricate structure than those of $\mathsf W_n^{1/2}(\mu,k)$. In the absence of suitable analytical methods, previous studies have relied entirely on high-precision numerical evaluations of the ABJM matrix integral \re{Z}. 

In this subsection, we apply the technique described above to compute $\mathsf W_n^{1/6}(\mu,k)$ starting from the relation \re{vev-GC} supplemented with the generating function \re{W-cf}. 
Since the two generating functions in \re{WW-gen} are related to each other, it is convenient to examine their difference. To this end we use the relation \re{W1/2-cf} to rewrite \re{W-cf} as
\begin{align}\label{W-minus-W} 
 \mathcal W_n(\mu,k)=\frac12 \mathcal W^{1/2}_n(\mu,k)+i \,\widetilde {\mathcal W}_n(\mu,k)\,.
 \end{align}
The function $\widetilde {\mathcal W}_n(\mu,k)$ is given by
\begin{align}\label{Wt}
 \widetilde {\mathcal W}_n(\mu,k)=\frac{ i^{n-1}}{\sin(\frac{2\pi n}{k})}\Big[ \mathcal B_n(z) \Phi'(\mu) + \mathcal C_n(z)  \lr{\log D(\mu)}' \Big],
\end{align}
where prime denotes a derivative with respect to the chemical potential $\mu=\log z$. Note that the ratio  $\mathcal W^{1/2}_n(\mu,k)/\widetilde {\mathcal W}_n(\mu,k)$ is a real function of the chemical potential. 

The difference function \re{Wt} satisfies the relation analogous to~\eqref{vev-GC}
\begin{align}\label{Wt-GC}
\widetilde{\mathcal W}_n(\mu,k) = \big\langle \widetilde{\mathsf W}_n(\mu,k) \big\rangle_{_{\rm GC}},
\end{align}
where $\widetilde{\mathsf W}_n(\mu,k)$ is related to the functions in~\eqref{WW-gen} through a relation similar to~\eqref{W-minus-W}. It is explicitly given by
\begin{align}\label{W-Re}
\widetilde{\mathsf W}_n(\mu,k)={i^{n-1} e^{2n\mu/k}\over k\, \sin({2\pi n\over k})}\Re\left[\mathsf W_n^{\rm pert}(\mu,k) +\mathsf W_n^{\rm np}(\mu,k) \right].
\end{align}
The perturbative function $\mathsf W_n^{\rm pert}(\mu,k)$ is defined in \re{W-pert}. The nonperturbative part $\mathsf W_n^{\rm np}(\mu,k)$ is given by the expansion \re{W-np} with coefficients that are to be determined. By construction, the function $\widetilde{\mathsf W}_n(\mu,k)$ depends on the real part of these coefficients, while their imaginary part can be derived from $\mathsf W^{1/2}_n(\mu,k)$.

According to the definition \re{A-n=1} and \re{A-n=2}, the functions $\mathcal B_n(z)$ and $\mathcal C_n(z)$ depend on the chemical potential through the functions $\Phi(\mu)$ and $D(\mu)$ given by \re{Phi-exp}. These functions depend on the effective chemical potential $\mu_{\rm eff}$ and the phase $\varphi(\mu_{\rm eff})$ introduced in \re{eq:mueff} and \re{varphi}, respectively. 

We have seen in the previous subsection that it is advantageous to redefine the phase according to \re{tildephi}. In terms of the modified phase $\widetilde\varphi$, the relation 
\re{Phi-exp} takes the form
\begin{align}\label{Phi-imp}
\Phi=\widetilde\varphi+\sum_{m\ge 1} \phi_{m0}(\widetilde \varphi) \,Q_{\rm w}^m \,,
\end{align}
where the dependence on the membrane parameter $Q_{\rm m}$ arises from the expansion of $\widetilde\varphi=\widetilde\varphi(\varphi,Q_{\rm m})$ in powers of $Q_{\rm m}$. Comparing the relations \re{Phi-imp} and \re{Phi-exp}, we find that the remaining expansion coefficients $\phi_{m\ell}(\varphi)$ in \re{Phi-exp} can be expressed in terms of  $\phi_{m0}(\varphi)$ and $\widetilde J_b(\mu_{\rm eff},k)=\sum \tilde b_\ell(k) Q_{\rm m}^\ell$.
 
Applying the relations \re{D-imp} and \re{Phi-imp}, we can derive from \re{Wt} an equivalent representation of $\widetilde {\mathcal W}_n(z,k)$
\begin{align}\label{tildeW-imp}
\widetilde {\mathcal W}_n(z,k)=\frac{ i^{n-1}}{\sin(\frac{2\pi n}{k})}{\partial \mu_{\rm eff}\over \partial \mu}\bigg[{\partial\widetilde\varphi\over\partial\mu_{\rm eff}}\widetilde{\mathcal B}_n (\widetilde\varphi,Q_{\rm w})+ {2\over k} \widetilde{\mathcal C}_n(\widetilde\varphi,Q_{\rm w})\bigg].
\end{align}
Importantly, the functions $\widetilde{\mathcal B}_n$ and $\widetilde{\mathcal C}_n$ do not have an explicit dependence on the membrane parameter $Q_{\rm m}$. They are given by linear combinations of the functions ${\mathcal B}_n$ and ${\mathcal C}_n$ introduced in \re{W-cf}
\begin{align}\notag\label{BCt}
{}& \widetilde{\mathcal B}_n(\widetilde\varphi,Q_{\rm w})={\partial\Phi\over\partial \widetilde\varphi}\Big(\mathcal B_n -(\cot\Phi -\partial_\Phi J_\gamma)\mathcal C_n \Big),
\\[2mm]
{}& \widetilde{\mathcal C}_n(\widetilde\varphi,Q_{\rm w})=\mathcal C_n\Big(1 -2Q_{\rm w}\partial_{Q_{\rm w}} J_\gamma \Big)-2 Q_{\rm w}{\partial\Phi \over \partial Q_{\rm w} }\Big(\mathcal B_n - (\cot\Phi -\partial_\Phi J_\gamma)\mathcal C_n\Big),
\end{align}
where the notation was introduced for  $ J_\gamma(\Phi,Q_{\rm w}) =\log\Big(1+\sum \gamma_{m0}(\Phi) Q_{\rm w}^m \Big)$, see \re{D-imp}.  
Details of the derivation of the relation \re{tildeW-imp} are presented in Appendix~\ref{appC}. 
 
The derivatives in \re{tildeW-imp} originate from rewriting the derivatives in~\eqref{Wt} in terms of $\mu_{\rm eff}$ and $\widetilde\varphi$. They can be evaluated using \re{eq:mueff} and \re{tildephi} as 
\begin{align}\notag
  {\partial \mu_{\rm eff}\over \partial \mu}=f_{\rm m}(Q_{\rm m}) 
\,,\qqqquad {\partial\widetilde\varphi\over\partial\mu_{\rm eff}}=- \frac{4}{\pi k}\mu_{\mathrm{eff}}+4V_{\rm m}(Q_{\rm m})\,,
\end{align}
where a shorthand notation was introduced for two functions depending on $Q_{\rm m}$ only
\begin{align}\notag\label{fm}
{}& f_{\rm m}=1-{2\over C} \sum_{\ell\ge 1} \ell\, a_\ell(k) Q_{\rm m}^\ell\,,
\\[2mm]
{}& V_{\rm m}=-{\pi\over 4} \widetilde J_b{}'(\mu_{\rm eff})={\pi\over 2} \sum_{\ell\ge 1} \ell\,\widetilde b_\ell(k) Q_{\rm m}^\ell\,.
\end{align}
The coefficients $a_\ell(k)$ were determined in Section~\ref{sect:mueff}, whereas the coefficients $\widetilde b_\ell(k)$ remain undetermined. 

As follows from the definition \re{Q-def}, the functions \re{fm} are invariant under $\mu\to \mu+i\pi$. As a consequence, they are not affected by the grand canonical average \re{vev-GC2}, since 
\begin{align}\label{pro1}
\vev{f(Q_{\rm m}) \mathsf U(\mu)}_{_{\rm GC}} =f(Q_{\rm m}) \vev{\mathsf U(\mu)}_{_{\rm GC}}
\end{align}
for an arbitrary function of $Q_{\rm m}$.
Taking this property into account, we can use the following ansatz for the function \re{W-Re}
 \begin{align}\label{Wt-ansatz} 
    \widetilde{\mathsf W}_n(\mu,k)=\frac{i^{n-1} e^{\frac{2n\mu_{\mathrm{eff}}}{k}}}{\sin\left(\frac{2\pi n}{k}\right)}f_{\rm m}\sum_{m,\ell=0 }^{\infty}
   \widetilde{\mathsf w}_{m,\ell}(\mu_{\mathrm{eff}})\,Q_{\rm w}^mQ_{\rm m}^\ell\,.
\end{align}
Replacing $\widetilde{\mathcal W}_n(\mu,k)$ and $\widetilde{\mathsf W}_n(\mu,k)$ in \re{Wt-GC} with their expressions \re{tildeW-imp} and \re{Wt-ansatz}, respectively, we 
arrive at the equation for the coefficients $\widetilde{\mathsf w}_{m,\ell}(\mu_{\mathrm{eff}})$  
\begin{align}\label{eq1/6}
\VEV{e^{\frac{2n\mu_{\mathrm{eff}}}{k}}\sum_{m,\ell=0 }^{\infty}
   \widetilde{\mathsf w}_{m,\ell}(\mu_{\mathrm{eff}})\,Q_{\rm w}^mQ_{\rm m}^\ell}_{^{\rm GC}} =  -{4\over k}\lr{\frac{\mu_{\mathrm{eff}}}{\pi}-k V_{\rm m}(Q_{\rm m})}\widetilde{\mathcal B}_n(\widetilde\varphi,Q_{\rm w}) + {2\over k} \widetilde{\mathcal C}_n(\widetilde\varphi,Q_{\rm w}) \,,
\end{align}
where the functions $\widetilde{\mathcal B}_n$ and $\widetilde{\mathcal C}_n$ are defined in \re{BCt}. This relation is similar to the analogous relation \re{eq1/2} for $1/2$ BPS Wilson loop. The important difference is that the equation \re{eq1/6} contains a term linear in $\mu_{\rm eff}$. As we show below, this leads to an important difference in the properties of its solution. 

\subsection{Solution to the master equation}
\label{eq:sec4.4}

In this subsection, we solve the master equation \re{eq1/6}, supplemented by the requirement that its solutions $\widetilde{\mathsf w}_{m,\ell}(\mu_{\mathrm{eff}})$ be independent of the phase $\varphi$.

Performing the grand canonical average \re{vev-GC2} on the left-hand side of \re{eq1/6}, we apply \re{osc1} and replace the phase $\varphi$ with its modified counterpart \re{varphi}. The resulting expression takes the form of a double series in $Q_{\rm w}$ and $Q_{\rm m}$ with the coefficients that depend on $\widetilde\varphi$ and $\widetilde{\mathsf w}_{m,\ell}$, which we treat as independent quantities. A nontrivial consistency requirement of the equation \re{eq1/6} is that both sides must exhibit the same functional dependence on $\widetilde\varphi$ at each order in $Q_{\rm w}$ and $Q_{\rm m}$. Imposing this condition allows us to determine the coefficient functions $\widetilde{\mathsf w}_{m,\ell}(\mu_{\mathrm{eff}})$ and to reproduce the previously obtained values of the grand-potential coefficients $\widetilde f_{m,\ell}$.
 
We start by computing the coefficient $\widetilde{\mathsf w}_{00}(\mu_{\mathrm{eff}})$ which governs the perturbative contribution to \re{Wt-ansatz}. For this purpose, it is sufficient to neglect the nonperturbative corrections to the functions $\widetilde{\mathcal B}_n$ and $\widetilde{\mathcal C}_n$. Using the expressions for ${\mathcal B}_n$ and ${\mathcal C}_n$ derived in \cite{Boldis:2025yll}, we obtain 
\begin{align}\notag
{}& \widetilde{\mathcal B}_n =-\frac14 e^{2n\mu_{\mathrm{eff}}/k}\lr{1+O(Q_{\rm w})}\,,
\\ 
{}& \widetilde{\mathcal C}_n =-\frac12 e^{2n\mu_{\mathrm{eff}}/k} \sum_{j=1}^n \cot\lr{2\pi j\over k}\lr{1+O(Q_{\rm w})}\,.
\end{align}
Within this approximation, the grand canonical averaging in \re{eq1/6} can be omitted 
(see \re{app-av}). As a result, setting 
$Q_{\rm w}=Q_{\rm m}=0$ on both sides of \eqref{eq1/6}, we obtain
\begin{align}\label{w00-1/6} 
\widetilde{\mathsf w}_{00}(\mu_{\mathrm{eff}})={1\over k} \bigg({\mu_{\mathrm{eff}}\over\pi}-\sum_{j=1}^n \cot\lr{2\pi j\over k}\bigg).
\end{align} 
This relation agrees with the perturbative part of $1/6$ BPS Wilson loop computed numerically in \cite{Okuyama:2016deu} and derived in \cite{Boldis:2025yll}.

Going beyond the perturbative approximation, we look for a solution to \re{eq1/6} in the form
\begin{align}\label{f-C}
\sum_{m,\ell=0 }^{\infty}
   \widetilde{\mathsf w}_{m,\ell}(\mu_{\mathrm{eff}})\,Q_{\rm w}^mQ_{\rm m}^\ell={1\over k}f_{\rm w} \lr{\frac{\mu_{\mathrm{eff}}}{\pi}-k V_{\rm m}+Y_{\rm w}}\,.
\end{align}
The expression on the right-hand side involves two functions, $f_{\rm w}$ and $Y_{\rm w}$, which remain to be determined. In general, both functions may depend on the parameters $Q_{\rm w}$ and $Q_{\rm m}$. Requiring that the limit $Q_{\rm w},Q_{\rm m}\to 0$ of \re{f-C} reproduces \re{w00-1/6} provides the necessary boundary condition. Since the function $V_{\rm m}$ vanishes in this limit (see \re{fm}), this requirement uniquely fixes the boundary values of $f_{\rm w}$ and $Y_{\rm w}$.

Substituting the ansatz \re{f-C} into \re{eq1/6} we find that the functions $f_{\rm w}$ and $Y_{\rm w}$ satisfy the relations
\begin{align}\notag\label{sys1/6}
{}& \VEV{e^{\frac{2n\mu_{\mathrm{eff}}}{k}} f_{\rm w}}_{_{\rm GC}} = -4 \widetilde{\mathcal B}_n(\widetilde\varphi,Q_{\rm w}) \,, 
\\[2mm]
{}& \VEV{e^{\frac{2n\mu_{\mathrm{eff}}}{k}} \lr{\frac{\mu_{\mathrm{eff}}}{\pi}+Y_{\rm w}}f_{\rm w}}_{_{\rm GC}} = -4{\frac{\mu_{\mathrm{eff}}}{\pi}}\widetilde{\mathcal B}_n(\widetilde\varphi,Q_{\rm w}) + 2\widetilde{\mathcal C}_n(\widetilde\varphi,Q_{\rm w}) \,.
\end{align} 
The first relation follows from comparing the coefficient of $\mu_{\rm eff}$ on both sides of \re{eq1/6}. To arrive at the second relation we used that $V_m$ depends on $Q_{\rm m}$ and applied the identity \re{pro1}. 

Since the right-hand side of \re{sys1/6} is independent of $Q_{\rm m}$, the functions $f_{\rm w}$ and $Y_{\rm w}$ can only depend on $Q_{\rm w}$. We can compute these functions up to order $O(Q_{\rm w})$ by noticing that the grand canonical average satisfies
$\vev{\mathsf U(\mu)}_{_{\rm GC}}=\mathsf U(\mu)(1+O(Q_{\rm w}^2))$ and, therefore,  $\vev{\mathsf U(\mu)}_{_{\rm GC}}$ coincides with $\mathsf U(\mu)$ at order $O(Q_{\rm w})$. Omitting the average in \re{sys1/6} and replacing the functions $ \widetilde{\mathcal B}_n$ and $\widetilde{\mathcal C}_n$ by their expressions,  we compute the leading $O(Q_{\rm w})$ corrections to $f_{\rm w}$ and $Y_{\rm w}$:
\begin{itemize}
\item
For $n=1$ we find
\begin{align}\notag\label{fY1}
{}& f_{\rm w} =1+2Q_{\rm w} +O(Q_{\rm w}^2)\,, 
\\[2mm]
{}& Y_{\rm w}= -\cot\lr{\ft{2\pi}{k}}\left[1-2Q_{\rm w}
+O(Q_{\rm w}^2)\right].
\end{align}
\item
For $n\ge 2$ we have instead
\begin{align}\notag\label{fY2}
{}&  f_{\rm w} =1- Q_{\rm w}{\sin^2({2\pi n\over k})\over \sin^2({2\pi\over k})}+O(Q_{\rm w}^2)\,,
\\
{}& Y_{\rm w} =-\sum_{j=1}^n \cot\big(\ft{2\pi j}{k}\big)+{\sin \left(\frac{2 \pi  n}{k}\right) \left(\cos
   ^2\left(\frac{2 \pi  n}{k}\right)-\cos \left(\frac{4 \pi }{k}\right)\right)\over \sin ^3\left(\frac{2 \pi }{k}\right) \sin \big(\frac{2 \pi  (n-1)}{k}\big)}Q_{\rm w} + O(Q_{\rm w}^2)\,.
\end{align}
\end{itemize}
Subleading corrections to these relations can be found in Appendix~\ref{appB}.

Combining together the relations \re{Wt-ansatz} and \re{f-C}, we arrive at the following result for $1/6$ BPS Wilson loop
\begin{align}\label{W1/6-fin} 
\widetilde{\mathsf W}_n(\mu,k)=\frac{i^{n-1} e^{\frac{2n\mu_{\mathrm{eff}}}{k}}}{k\sin\left(\frac{2\pi n}{k}\right)}f_{\rm m}\,f_{\rm w}\left[\frac{\mu_{\mathrm{eff}}}{\pi}-k V_{\rm m}+Y_{\rm w}\right].
\end{align}
It involves four different functions depending on the nonperturbative parameters $Q_{\rm m}$ and $Q_{\rm w}$. 

\paragraph{The world-sheet functions} 
 $f_{\rm w}$ and $Y_{\rm w}$ depend on  the world-sheet parameter $Q_{\rm w}$ and the winding number $n$. Their explicit expressions are given by \re{fY1} and \re{fY2}. 

\paragraph{The membrane functions} 
$f_{\rm m}$ and $V_{\rm m}$ depend on the membrane parameter $Q_{\rm m}$ and are independent of the winding number $n$. According to the definition \re{fm}, these functions depend on the coefficients $a_\ell(k)$ and $\widetilde b_\ell(k)$ parameterizing the nonperturbative corrections to the grand potential \re{J-gen}. Replacing the former coefficients with their expressions \re{aa}, we find
\begin{align}\notag
f_{\rm m} {}& =1+4\cos \left(\ft{\pi  k}{2}\right)Q_{\rm m}+4 (5 \cos (\pi  k)+4)Q_{\rm m}^2
\\[2mm]
{}& +8\cos
   \left(\ft{\pi  k}{2}\right) (28 \cos (\pi  k)+3 \cos (2 \pi 
   k)+19) Q_{\rm m}^3 +O\left(Q^4\right).
\end{align}
This relation holds for arbitrary values of $k$. For odd $k$, the coefficients of odd powers of $Q_{\rm m}$ vanish so that $f_{\rm m}$ is even in $Q_{\rm m}$.

As mentioned above, the coefficients $\widetilde b_\ell(k)$ and, consequently, the function $V_{\rm m}$ defined in \re{fm} remain undetermined. In close analogy with the grand potential \cite{Hatsuda:2013gj}, these coefficients can be fixed by requiring that the Wilson loop \re{W1/6-fin} is a well-defined function of the Chern-Simons level $k$.

An important observation is that the world-sheet function $Y_{\rm w}$, defined in \re{fY1} and \re{fY2}, develops poles for certain values of $k$, which are specified below. Recall that the ABJM matrix integral \re{Z} is well defined for $k>2n$; it is therefore not surprising that $Y_{\rm w}$ exhibits poles for $k\leq 2n$. However, at higher orders in $Q_{\rm w}$, additional poles appear for rational values of $k>2n$.
For the Wilson loop \re{W1/6-fin} to be well defined, these singularities must
cancel in the linear combination $(-k\,V_{\rm m}+Y_{\rm w})$ against the
corresponding poles of the membrane contribution $V_{\rm m}$.
This requirement provides a strong constraint on the function $V_{\rm m}$.

Expanding the world-sheet function as
$
Y_{\rm w}=\sum_{m\geq 1} y_m(k,n)\,Q_{\rm w}^m
$,
we obtain
\begin{align}\label{poles}
-k\,V_{\rm m}+Y_{\rm w}
= -\frac{\pi k}{2}\sum_{\ell\geq 1} \ell\,\widetilde b_\ell(k)\,e^{-2\mu \ell}
+\sum_{m\geq 1} y_m(k,n)\,e^{-4m\mu/k}\,.
\end{align}
As argued in \cite{Hatsuda:2013gj,Okuyama:2016deu}, the coefficients $\widetilde b_\ell(k)$ and
$y_m(k,n)$ develop poles whenever the two exponential factors coincide, $e^{-2\ell\mu}=e^{-4m\mu/k}$.
This occurs for rational values of $k=2m/\ell$. Imposing the condition $k>2n$ further restricts the integers to
$m>n\ell$.
Requiring the cancellation of these spurious poles in
\eqref{poles} constrains the behavior of $\widetilde b_\ell(k)$ in the
vicinity of $k=2m/\ell$ 
\begin{align}\label{poles-can}
\widetilde b_\ell(k)
\sim
\frac{r_{m\ell}}{\pi m\,\bigl(k-2m/\ell\bigr)}\,,
\end{align}
where
$
r_{m\ell}=\res_{k=2m/\ell} y_m(k,n)
$
denotes the residue of $y_m(k,n)$ at the spurious pole for $m>n\ell$.

Notice that, in contrast to the coefficients $y_m(k,n)$, the functions
$\widetilde b_\ell(k)$ are independent of the winding number $n$.
The condition \re{poles-can} therefore implies that the residues
$r_{m\ell}$ must also be independent of $n$.
We have explicitly verified this property using the expressions for the
world-sheet function $Y_{\rm w}$ for $1\leq n\leq 5$ presented in
Appendix~\ref{appB} and in the ancillary Mathematica file.
Moreover, we have checked that the conjectured expressions \re{ab} for the
coefficients $\widetilde b_\ell(k)$ indeed satisfy the constraint
\re{poles-can}.
 
Conversely, one may follow the strategy of \cite{Hatsuda:2013gj} and bootstrap the
coefficients $\widetilde b_\ell(k)$ by imposing \re{poles-can}, together
with the explicit results for the functions $Y_{\rm w}$ collected in Appendix~\ref{appB}.
The resulting expression for the membrane function $V_{\rm m}$ reads
\begin{align}\notag\label{Vm}
V_{\rm m} {}& 
=\cos \left(\ft{\pi  k}{2}\right) \cot \left(\ft{\pi  k}{2}\right)Q_{\rm m}+\cot (\pi  k)\lr{4 +5\cos (\pi  k)}Q_{\rm m}^2
\\[2mm]
{}& +2\cos \left(\ft{\pi  k}{2}\right) \cot \left(\ft{3\pi  k}{2}\right)\lr{13 +19\cos (\pi  k)+9\cos (2\pi  k)}Q_{\rm m}^3+O(Q_{\rm m}^4)\,.
\end{align}
In the refined topological string description, this function  is related to NS free energy on local $\mathbb{P}^1\times \mathbb{P}^1$. Furthermore, it was proposed in \cite{Okuyama:2016deu} that for $n=1$ the expression inside the brackets in \re{W1/6-fin} is related to the quantum volume of the associated phase space.

Comparing the obtained expressions \re{lead} and \re{W1/6-fin}, we observe that, in distinction to $1/2$ BPS Wilson loop, it less supersymmetric cousin,  $1/6$ BPS Wilson loop, receives the contribution from membrane instantons. Interestingly, it is described by two  universal functions $f_{\rm m}$ and $V_{\rm m}$ which are independent of the winding number $n$. 

We verified that the relation \re{W1/6-fin} is in agreement with the expressions for $1/6$ BPS Wilson loop obtained in \cite{Okuyama:2016deu} from high-precision numerical calculations

\section{Conclusions}
\label{sec:5}

Supersymmetric localization reduces the computation of a wide class of protected observables in ABJM theory to finite-dimensional matrix integrals. Due to their intricate structure, extracting their large-$N$ behavior for arbitrary values of the Chern–Simons level $k$ remains a highly nontrivial task. In this work, building on the techniques introduced in~\cite{Boldis:2025yll}, we developed a bootstrap framework within the Fermi-gas approach that allows for a systematic computation of instanton corrections to the grand potential and to multiply wound supersymmetric Wilson loops.

Nonperturbative corrections in ABJM theory were extensively studied using a variety of complementary approaches, including refined topological string descriptions~\cite{Nosaka:2015iiw,Hatsuda:2012dt,Hatsuda:2012hm,Hatsuda:2013oxa} and direct computations of quantum M2-brane contributions~\cite{Giombi:2023vzu,Gautason:2023igo,Beccaria:2023ujc,Gautason:2025per,Gautason:2025plx}. In parallel, high-precision numerical analyses provided strong support for these results and motivated a number of conjectural expressions~\cite{Hatsuda:2013yua,Okuyama:2016deu,Hatsuda:2015gca}. Nevertheless, a first-principle derivation of the nonperturbative corrections directly from the underlying matrix-model formulation  remained elusive.

The bootstrap framework developed in this work relies solely on ABJM matrix model representation. By exploiting exact functional relations and consistency conditions satisfied by grand-canonical observables in the Fermi gas formulation, we provided analytical derivations of several relations for the free energy and supersymmetric Wilson loop previously known only conjecturally, either from refined topological string arguments or from high-precision numerical analyses. Our results thus provide further evidence for the intricate web of dualities underlying ABJM theory.

We found that the nonperturbative corrections to $1/2$ and $1/6$ BPS Wilson loops,  expressed as functions of the modified chemical potential $\mu_{\rm eff}$, exhibit qualitatively different structures. While the $1/2$ BPS loop receives contributions exclusively from world-sheet instantons (see footnote ${}^{\ref{foot}}$),  the $1/6$ BPS Wilson loop also involves genuine membrane-instanton effects. Remarkably, for arbitrary winding number, these membrane contributions are governed by two universal functions that are independent of the winding number. Our results thus provide an analytic realization of the general ansatz for $1/2$ BPS Wilson loop proposed in~\cite{Hatsuda:2013yua}, and elucidate the origin of several features of $1/6$ BPS Wilson loop previously observed numerically in~\cite{Okuyama:2016deu}.  In the M-theory description, it was argued in \cite{Drukker:2008zx,Rey:2008bh,Giombi:2023vzu} that the difference between $1/2$ and $1/6$ BPS Wilson loops originates from an extra smearing of the relevant M2 brane solution over a $\mathbb{CP}^1$.

Despite these advances, several open questions remain. Within the framework described above, we were able to determine the general structure of the grand potential~\re{J-eff} and to derive analytic expressions for the world-sheet instanton coefficients $d_m(k)$ in~\eqref{f1f2f3}, as well as for the membrane-instanton coefficients $\tilde b_\ell(k)$ in~\eqref{JM2}. However, the membrane coefficients $\tilde c_\ell(k)$ remain undetermined, since they do not contribute to the Wilson loops considered in this work. These coefficients were conjectured in~\cite{Hatsuda:2013gj} to be related to $\tilde b_\ell(k)$. Establishing this relation directly from the matrix-model formulation is an important open problem.

Another important issue concerns the behavior of the supersymmetric Wilson loops at special values of the Chern–Simons level $k$, where the localization expressions develop singularities while the corresponding holographic predictions remain finite. In particular, the matrix-model integral \re{Z} is well defined only for levels satisfying $k>2n$. At $k=2n$, the integration at infinity in \re{Z} produces a pole. The same pole appears in the supersymmetric Wilson loops, originating from the factor $1/\sin(2\pi n/k)$ in \re{W-ans}. In the large-$N$ limit and fixed $k$, $1/2$ BPS Wilson loop admits a dual M-theory description in terms of an M2-brane wrapping the M-theory circle. It was shown in~\cite{Giombi:2023vzu} that, for $n=1$ and $k>2$, the same factor of $1/\sin(2\pi/k)$ is precisely reproduced by the one-loop contribution to the wrapped M2-brane partition function. Interestingly, for $k=1$ and $k=2$ the holographic prediction remains finite, whereas the localization result becomes singular. Understanding the origin of this mismatch remains an important open problem.

It would also be interesting to generalize our approach to other observables, including the latitude Wilson loop in ABJM theory~\cite{Griguolo:2021rke,Bianchi:2018bke}, and to Wilson loops in the presence of deformations, such as real masses~\cite{Armanini:2024kww,Nosaka:2015iiw} or the squashing of $S^3$. These deformations break conformal invariance while preserving part of the supersymmetry. Notably, it has been conjectured (see~\cite{Bobev:2025ltz} and references therein) that the large-$N$ partition function still admits an Airy-function representation, suggesting that Wilson loops may admit an analogous structure.

Finally, we note a suggestive similarity between the structure of nonperturbative corrections in ABJM theory and those in four-dimensional superconformal Yang--Mills theories. In the latter, various observables can be computed exactly in terms of Fredholm determinants of specific integrable operators, allowing for a systematic analysis of exponentially suppressed corrections (see \cite{Beccaria:2022ypy,Bajnok:2024epf,Bajnok:2024ymr} and references therein). From this perspective, the instanton contributions in ABJM theory may be viewed as an intrinsically three-dimensional counterpart of nonperturbative effects controlled by integrable operators in four-dimensions. It would be interesting to exploit this analogy further, with the aim of identifying and characterizing an analog of membrane-instanton corrections in four-dimensional gauge theory. Conversely, identifying observables in ABJM theory that admit a Fredholm-determinant representation, could provide a unified description of nonperturbative effects across dimensions and illuminate the role of integrability in strongly coupled gauge theories.

\section*{Acknowledgements} 

We would like to thank Marco Billo',  Luca Griguolo, Alberto Lerda, Kazumi Okayama and Arkady Tseytlin for interesting discussions. We are also grateful to Marcos Mari\~no for his valuable comments and critical remarks which improved the presentation and completeness of the paper. 
The research of B.B. was supported by the Doctoral Excellence Fellowship Programme funded by the National
Research Development and Innovation Fund of the Ministry of Culture and Innovation and the Budapest
University of Technology and Economics, under a grant agreement with the National Research, Development
and Innovation Office (NKFIH). The work of G.K. and A.T.  was supported by the French National Agency for Research grant ``Observables'' (ANR-24-CE31-7996).

\appendix

\section{Grand canonical ensemble averages}\label{appA}

In this appendix, we describe the properties of the average in the grand canonical ensemble defined in \re{vev-GC2}.
As an example, we choose the function $\mathsf U(\mu)$ of the form
\begin{align}\label{app-U}
\mathsf U(\mu)=\sum_{m,\ell \ge 0} \mathsf u_{m\ell}(\mu_{\rm eff})\,Q_{\rm w}^mQ_{\rm m}^\ell\,,
\end{align}
where the effective chemical potential is defined in \re{eq:mueff} and the parameters $Q_{\rm w}$ and $Q_{\rm m}$ are introduced in \re{Q-def}.

It follows from \re{eq:mueff} that a shift of the chemical potential $\mu\to\mu+i\pi$ induces a corresponding shift $\mu_{\rm eff}\to\mu_{\rm eff}+i\pi$. As a consequence, for integer $m$ we find
\begin{align}
Q_{\rm w}(\mu+2i\pi m) = e^{4i \pi m/k}  Q_{\rm w}(\mu)\,,\qqqquad Q_{\rm w}(\mu+2i\pi m)=Q_{\rm w}(\mu). 
\end{align}
The second relation implies that the average of any function depending only on $Q_{\rm w}(\mu)$ coincides with the function itself.

Furthermore, using the relations \re{osc1} and \re{tildephi} we obtain
\begin{align}\label{J-J}  
e^{J(\mu+2im\pi,k)-J(\mu,k)}=Q_{\rm w}^{2m^2} \exp\lr{2im\lr{{\pi\over 2}- \widetilde\varphi(\mu_{\rm eff})}-{4i\pi\over 3k}m(4m^2-1) +O(Q_{\rm w})}\,.
\end{align}
Due to the first factor on the right-hand of \re{J-J}, the contribution of various terms in the numerator and denominator of \re{vev-GC2} is suppressed by the factor of $Q_{\rm w}^{2m^2}$. 

As a consequence, the leading contribution to $\vev{\mathsf U(\mu)}_{{\rm GC}}$ comes from terms with $m=0$ and $m=\pm 1$
\begin{align}\notag\label{U-exp}
  \vev{\mathsf U(\mu)}_{_{\rm GC}} =\mathsf U(\mu){}&+(\mathsf U(\mu+ 2\pi i)-\mathsf U(\mu))e^{J(\mu + 2\pi i)-J(\mu)}
\\[2mm]
{}&+(\mathsf U(\mu-2\pi i)-\mathsf U(\mu))e^{J(\mu -2\pi i)-J(\mu)}+O(Q_{\rm w}^{4})\,.
\end{align}
We deduce from this relation that the grand canonical average generates nonperturbative corrections 
\begin{align}\label{app-av}
  \vev{\mathsf U(\mu)}_{_{\rm GC}} =\mathsf U(\mu) + O(Q_{\rm w}^2)\,.
\end{align}
Note that $O(Q_{\rm w})$ terms in $\mathsf U(\mu)$ are not affected by the average.

Substituting \re{app-U} into \re{U-exp}, we find that the grand canonical average  $\mathcal U(\mu) = \vev{\mathsf U(\mu)}_{_{\rm GC}} $ admits an expansion of the same form as \re{app-U}
\begin{align}
\mathcal U(\mu) = \vev{\mathsf U(\mu)}_{_{\rm GC}} = \sum_{m,\ell \ge 0} u_{m\ell}(\mu_{\rm eff})\,Q_{\rm w}^mQ_{\rm m}^\ell\,.
\end{align}
The coefficients $u_{m\ell}(\mu_{\rm eff})$ are given by linear combinations of the $\mathsf u-$coefficients. In particular, for $\ell\ge 0$ we find
\begin{align}\notag
u_{0\ell}{}&=\mathsf u_{0\ell}(\mu_{\rm eff})\,,  \qquad u_{1\ell}=\mathsf u_{1\ell}(\mu_{\rm eff})\,,
\\[2mm]
u_{2\ell}{}&=\mathsf u_{2\ell}(\mu_{\rm eff})+\left[ e^{\, 2i\widetilde\varphi(\mu_{\rm eff})+{4i\pi\over k}}\lr{\mathsf u_{0\ell}(\mu_{\rm eff})-\mathsf u_{0\ell}(\mu_{\rm eff}-2i\pi)} + {\rm c.c.} \right].
\end{align}
Note that the $u-$coefficients acquire a dependence on the modified phase $\widetilde\varphi(\mu_{\rm eff})$ 
defined in \re{tildephi} and \re{varphi}. As a result, they become rapidly oscillating functions at large $\mu_{\rm eff}$.

\section{Derivation of \re{tildeW-imp}}\label{appC}

The relation \re{tildeW-imp} provides the generating function of the $1/6$ BPS Wilson loop,
$\widetilde{\mathcal W}_n(\mu,k)$, in terms of the functions defined in \re{BCt}. It can be
obtained from \re{Wt} by substituting the expressions for the derivatives $\Phi'(\mu)$ and
$\lr{\log D(\mu)}'$.

According to \re{Phi-imp}, the phase $\Phi(\mu)$ depends on the chemical potential through
the two functions $Q_{\rm w}(\mu_{\rm eff})$ and $\widetilde\varphi(\mu_{\rm eff})$, defined
in \re{Q-def} and \re{tildephi}, respectively. This allows us to express its derivative as
\begin{align}\label{der-Phi}
\Phi'(\mu) = \frac{\partial \mu_{\rm eff}}{\partial \mu}\left[\frac{\partial\widetilde\varphi}{\partial \mu_{\rm eff}}\,\frac{\partial\Phi}{\partial \widetilde\varphi}- \frac{4Q_{\rm w}}{k}
\frac{\partial\Phi}{\partial Q_{\rm w}}\right],
\end{align}
where we used $\partial Q_{\rm w} /\partial {\mu_{\rm eff}}=-4 Q_{\rm w} /k$. 

Furthermore, using \re{D-imp}, we rewrite $\log D(\mu)$ as
\begin{align}\notag
{}& \log D(\mu) = \frac{2\mu_{\rm eff}}{k}
- \log\bigl(2\sin\Phi\bigr)
+ J_\gamma(\Phi,Q_{\rm w})\,, \\[1.5mm]
{}& J_\gamma(\Phi,Q_{\rm w})
= \log \Bigl(1+\sum_{m\ge 1}\gamma_{m0}(\Phi) Q_{\rm w}^m \Bigr).
\end{align}
In complete analogy with \re{der-Phi}, its derivative is given by
\begin{align}
(\log D(\mu))'
=  -\bigl(\cot\Phi - \partial_\Phi J_\gamma(\Phi,Q_{\rm w})\bigr)\,\Phi'
+ \frac{\partial \mu_{\rm eff}}{\partial \mu}\Bigl(\frac{2}{k} - \frac{4Q_{\rm w}}{k}\,\partial_{Q_{\rm w}} J_\gamma(\Phi,Q_{\rm w})\Bigr).
\end{align}
Taking these relations into account, we obtain
\begin{align}
\mathcal B_n \,\Phi' + \mathcal C_n \,(\log D)'
= \frac{\partial\widetilde\varphi}{\partial\mu_{\rm eff}}\,\widetilde{\mathcal B}_n
+ \frac{2}{k}\,\widetilde{\mathcal C}_n\,,
\end{align}
where the functions $\widetilde{\mathcal B}_n$ and $\widetilde{\mathcal C}_n$ are defined
in \re{BCt}. Note that, similarly to the phase $\Phi$, these functions depend on $\mu$ only
through $Q_{\rm w}(\mu_{\rm eff})$ and $\widetilde\varphi(\mu_{\rm eff})$.

\section{World-sheet coefficient functions of $1/6$ BPS Wilson loop}\label{appB}

In this appendix, we present expressions for the coefficient functions $f_{\rm w}$ and $Y_{\rm w}$ entering $1/6$ BPS Wilson loop \re{W1/6-fin}. They were obtained by solving the equations \re{sys1/6} for different values of the winding number $n$.

\subsection*{$\bm{n=1}$}

\newcommand {\blue}{\textcolor{blue}} 
\newcommand {\Q}{Q_{\rm w}}

\begin{align} 
f_{\rm w}{}& =1+2\blue\Q-\blue{\Q^2}+2\blue{\Q^3}-7\blue{\Q^4}+16 \cos ^2\left(\ft{2 \pi }{k}\right) \left(3-\cos \left(\ft{4 \pi }{k}\right)\right)\blue{\Q^5}+\mathcal O\left(\blue{\Q^6}\right)
\\[2mm] \notag
Y_{\rm w} {}&= -\cot \left(\ft{2 \pi }{k}\right)+2 \cot \left(\ft{2 \pi }{k}\right) \blue\Q -2
   \cot \left(\ft{4 \pi }{k}\right) \blue{\Q^2}
   -4 \left(5 \cos \left(\ft{2 \pi
   }{k}\right)+2 \cos \left(\ft{6 \pi }{k}\right)\right) \csc \left(\ft{6 \pi
   }{k}\right)\blue{\Q^3}
\\[2mm]{}&   \notag
   +2 \left(80 \cos \left(\ft{4 \pi }{k}\right)+39 \cos
   \left(\ft{8 \pi }{k}\right)+8 \left(\cos \left(\ft{12 \pi
   }{k}\right)+5\right)\right) \csc \left(\ft{8 \pi }{k}\right) \blue{\Q^4}
\\[2mm]{}&   
   -2 \left(557
   \cos \left(\ft{2 \pi }{k}\right)+492 \cos \left(\ft{6 \pi }{k}\right)+291 \cos
   \left(\ft{10 \pi }{k}\right)+92 \cos \left(\ft{14 \pi }{k}\right)+27 \cos
   \left(\ft{18 \pi }{k}\right)\right) \csc \left(\ft{10 \pi }{k}\right)
   \blue{\Q^5}+O\left(\blue{\Q^6}\right),
 \end{align}

\subsection*{$\bm{n=2}$}

\begin{align}\notag
    f_{\rm w}{}&=1-4 \cos ^2\left(\ft{2 \pi }{k}\right)\blue{\Q}+\left(4 \cos \left(\ft{4 \pi }{k}\right)+2\right)\blue{\Q^2}
\\[2mm] \notag
    &-8 \left(\cos \left(\ft{4 \pi }{k}\right)+1\right)\blue{\Q^3}+\left(40 \cos \left(\ft{4 \pi }{k}\right)+6 \cos \left(\ft{8 \pi }{k}\right)+33\right)\blue{\Q^4}
\\[2mm] {}&    
    -16\cos^2\lr{\ft{2\pi}{k}} \left(12\cos \left(\ft{4 \pi }{k}\right)+4\cos \left(\ft{8 \pi }{k}\right) +13\right)\blue{\Q^5}+ O(\blue{\Q^6}),
\\[2mm] \notag
Y_{\rm w}{}&=  
   -\cot \left(\ft{2 \pi }{k}\right)-\cot \left(\ft{4
   \pi }{k}\right)-2 \left(\cos \left(\ft{2 \pi }{k}\right)+\cos
   \left(\ft{6 \pi }{k}\right)\right) \csc \left(\ft{2 \pi }{k}\right)
   \blue{\Q}
\\[2mm] {}&  \notag  
   -\cos \left(\ft{4 \pi }{k}\right) \left(2 \cos \left(\ft{4 \pi
   }{k}\right)+1\right)^2 \csc \left(\ft{2 \pi }{k}\right) \sec \left(\ft{2 \pi
   }{k}\right) \blue{\Q^2}
\\[2mm] {}&    \notag
   -2 \left(19 \cos \left(\ft{2 \pi }{k}\right)+13 \cos
   \left(\ft{6 \pi }{k}\right)+9 \cos \left(\ft{10 \pi }{k}\right)+2 \cos
   \left(\ft{14 \pi }{k}\right)+\cos \left(\ft{18 \pi }{k}\right)\right) \csc
   \left(\ft{6 \pi }{k}\right)\blue{\Q^3}
\\[2mm] {}&    \notag
   -2 \left(28 \cos \left(\ft{4 \pi
   }{k}\right)+20 \cos \left(\ft{8 \pi }{k}\right)+10 \cos \left(\ft{12 \pi
   }{k}\right)+7 \cos \left(\ft{16 \pi }{k}\right)+2 \cos \left(\ft{20 \pi
   }{k}\right)+\cos \left(\ft{24 \pi }{k}\right)+13\right) \csc \left(\ft{8 \pi
   }{k}\right) \blue{\Q^4}
\\[2mm] {}&    \notag
   -2 \big(250 \cos \left(\ft{2 \pi }{k}\right)+214 \cos
   \left(\ft{6 \pi }{k}\right)+157 \cos \left(\ft{10 \pi }{k}\right)+102 \cos
   \left(\ft{14 \pi }{k}\right)+44 \cos \left(\ft{18 \pi }{k}\right)+24 \cos
   \left(\ft{22 \pi }{k}\right)
\\[2mm] {}&  
   +2 \cos \left(\ft{26 \pi }{k}\right)+\cos
   \left(\ft{30 \pi }{k}\right)\big) \csc \left(\ft{10 \pi }{k}\right)
   \blue{\Q^5}+O\left(\blue{\Q^6}\right),
\end{align}

\subsection*{$\bm{n=3}$}

\begin{align}\notag
    f_{\rm w}{}&=1-\left(2 \cos \left(\ft{4 \pi }{k}\right)+1\right)^2\blue{\Q}+3 \left(2 \cos \left(\ft{4 \pi }{k}\right)+1\right)^2\blue{\Q^2} 
\\[2mm]\notag
    &-\left(36 \cos \left(\ft{4 \pi }{k}\right)+18 \cos \left(\ft{8 \pi }{k}\right)+25\right)\blue{\Q^3}
\\[2mm]
    &+12 \left(\cos \left(\ft{4 \pi }{k}\right)+2\right) \left(2 \cos \left(\ft{4 \pi }{k}\right)+1\right)^2\blue{\Q^4}+\mathcal O\left(\blue{\Q^5}\right),
\\[2mm]\notag
Y_{\rm w}{}&=   
  -\cot \left(\ft{2 \pi }{k}\right)-\cot \left(\ft{4
   \pi }{k}\right)-\cot \left(\ft{6 \pi }{k}\right) 
   -\left(8 \cos \left(\ft{4
   \pi }{k}\right)+6 \cos \left(\ft{8 \pi }{k}\right)+2 \cos \left(\ft{12 \pi
   }{k}\right)+5\right) \csc \left(\ft{4 \pi }{k}\right) \blue{\Q}
\\[2mm]{}&    \notag
   -\left(38 \cos
   \left(\ft{4 \pi }{k}\right)+28 \cos \left(\ft{8 \pi }{k}\right)+18 \cos
   \left(\ft{12 \pi }{k}\right)+10 \cos \left(\ft{16 \pi }{k}\right)+2 \cos
   \left(\ft{20 \pi }{k}\right)+21\right) \csc \left(\ft{4 \pi }{k}\right)
   \blue{\Q^2}
\\[2mm]{}&    \notag
   -\ft{1}{2} \big(534 \cos \left(\ft{4 \pi }{k}\right)+454 \cos
   \left(\ft{8 \pi }{k}\right)+340 \cos \left(\ft{12 \pi }{k}\right)+220 \cos
   \left(\ft{16 \pi }{k}\right)+120 \cos \left(\ft{20 \pi }{k}\right)
 \\[2mm]{}&  \notag
   +54 \cos
   \left(\ft{24 \pi }{k}\right)+16 \cos \left(\ft{28 \pi }{k}\right)+2 \cos
   \left(\ft{32 \pi }{k}\right)+281\big) \csc \left(\ft{6 \pi }{k}\right) \sec
   \left(\ft{2 \pi }{k}\right) \blue{\Q^3}
\\[2mm]{}&  \notag  
   -2 \big(905 \cos \left(\ft{4 \pi
   }{k}\right)+799 \cos \left(\ft{8 \pi }{k}\right)+640 \cos \left(\ft{12 \pi
   }{k}\right)+464 \cos \left(\ft{16 \pi }{k}\right)+306 \cos \left(\ft{20 \pi
   }{k}\right)+173 \cos \left(\ft{24 \pi }{k}\right)
\\[2mm]{}&   
   +83 \cos \left(\ft{28 \pi
   }{k}\right)+34 \cos \left(\ft{32 \pi }{k}\right)+9 \cos \left(\ft{36 \pi
   }{k}\right)+\cos \left(\ft{40 \pi }{k}\right)+474\big) \csc \left(\ft{8 \pi
   }{k}\right)\blue{\Q^4}+O\left(\blue{\Q^5}\right),
\end{align}
 
\subsection*{$\bm{n=4}$}

\begin{align}\notag
    f_{\rm w}&=1-4 \left(\cos \left(\ft{2 \pi }{k}\right)+\cos \left(\ft{6 \pi }{k}\right)\right)^2\blue{\Q}+4 \cos ^2\left(\ft{4 \pi }{k}\right) \left(8 \cos \left(\ft{4 \pi }{k}\right)+2 \cos \left(\ft{8 \pi }{k}\right)+7\right)\blue{\Q^2}
    \\[2mm]\notag
    &-8 \left(2 \cos \left(\ft{4 \pi }{k}\right)+5\right) \left(\cos \left(\ft{2 \pi }{k}\right)+\cos \left(\ft{6 \pi }{k}\right)\right)^2\blue{\Q^3} 
    \\[2mm]
    &+\left(4 \left(84 \cos \left(\ft{4 \pi }{k}\right)+57 \cos \left(\ft{8 \pi }{k}\right)+28 \cos \left(\ft{12 \pi }{k}\right)+50\right)+26 \cos \left(\ft{16 \pi }{k}\right)\right)\blue{\Q^4} +\mathcal O\left(\blue{\Q^5} \right),
\\[2mm]\notag
Y_{\rm w}{}&=    
   -\cot \left(\ft{2 \pi
   }{k}\right)-\cot \left(\ft{4 \pi }{k}\right)-\cot \left(\ft{6 \pi }{k}\right)-\cot
   \left(\ft{8 \pi }{k}\right) 
\\[2mm]{}&    \notag
   -2 \left(10 \cos \left(\ft{2 \pi
   }{k}\right)+8 \cos \left(\ft{6 \pi }{k}\right)+6 \cos \left(\ft{10 \pi
   }{k}\right)+3 \cos \left(\ft{14 \pi }{k}\right)+\cos \left(\ft{18 \pi
   }{k}\right)\right) \csc \left(\ft{6 \pi }{k}\right) \blue{\Q}
\\[2mm]{}&    \notag
   -2 \big(53 \cos
   \left(\ft{4 \pi }{k}\right)+45 \cos \left(\ft{8 \pi }{k}\right)+32 \cos
   \left(\ft{12 \pi }{k}\right)+22 \cos \left(\ft{16 \pi }{k}\right)+12 \cos
   \left(\ft{20 \pi }{k}\right)+5 \cos \left(\ft{24 \pi }{k}\right)
\\[2mm]{}&   \notag
   +\cos
   \left(\ft{28 \pi }{k}\right)+28\big) \csc \left(\ft{4 \pi }{k}\right)
   \blue{\Q^2} 
   -2 \big(262 \cos \left(\ft{2 \pi }{k}\right)+241 \cos \left(\ft{6 \pi
   }{k}\right)+205 \cos \left(\ft{10 \pi }{k}\right)+160 \cos \left(\ft{14 \pi
   }{k}\right)
\\[2mm]{}&   \notag
   +110 \cos \left(\ft{18 \pi }{k}\right)+69 \cos \left(\ft{22 \pi
   }{k}\right)+40 \cos \left(\ft{26 \pi }{k}\right)+18 \cos \left(\ft{30 \pi
   }{k}\right)+6 \cos \left(\ft{34 \pi }{k}\right)+\cos \left(\ft{38 \pi
   }{k}\right)\big) \csc \left(\ft{2 \pi }{k}\right)\blue{\Q^3}
\\[2mm]{}&    \notag
   -\ft{1}{2}
   \big(28766 \cos \left(\ft{4 \pi }{k}\right)+26381 \cos \left(\ft{8 \pi
   }{k}\right)+22820 \cos \left(\ft{12 \pi }{k}\right)+18577 \cos \left(\ft{16 \pi
   }{k}\right)+14195 \cos \left(\ft{20 \pi }{k}\right)
\\[2mm]{}&   \notag
   +10133 \cos \left(\ft{24 \pi
   }{k}\right)+6723 \cos \left(\ft{28 \pi }{k}\right)+4109 \cos \left(\ft{32 \pi
   }{k}\right)+2291 \cos \left(\ft{36 \pi }{k}\right)+1145 \cos \left(\ft{40 \pi
   }{k}\right)   +499 \cos \left(\ft{44 \pi }{k}\right)
\\[2mm]{}&   
+181 \cos \left(\ft{48 \pi
   }{k}\right)+51 \cos \left(\ft{52 \pi }{k}\right)+10 \cos \left(\ft{56 \pi
   }{k}\right)+\cos \left(\ft{60 \pi }{k}\right)+14801\big) \csc \left(\ft{6 \pi
   }{k}\right) \sec \left(\ft{2 \pi }{k}\right) \sec \left(\ft{4 \pi
   }{k}\right) \blue{\Q^4}+O\left(\blue{\Q^5}\right), 
\end{align}
 
\subsection*{$\bm{n=5}$}

\begin{align}\notag
    f_{\rm w}&=1-\left(2 \cos \left(\ft{4 \pi }{k}\right)+2 \cos \left(\ft{8 \pi }{k}\right)+1\right)^2\blue\Q+2 \left(\cos \left(\ft{8 \pi }{k}\right)+2\right) \left(2 \cos \left(\ft{4 \pi }{k}\right)+2 \cos \left(\ft{8 \pi }{k}\right)+1\right)^2\blue{\Q^2}
    \\[2mm]\notag
    &-\left(2 \cos \left(\ft{4 \pi }{k}\right)+2 \cos \left(\ft{8 \pi }{k}\right)+1\right)^2 \left(4 \cos \left(\ft{4 \pi }{k}\right)+6 \cos \left(\ft{8 \pi }{k}\right)+13\right)\blue{\Q^3} 
    \\[2mm]
    &+2 \left(2 \cos \left(\ft{4 \pi }{k}\right)+2 \cos \left(\ft{8 \pi }{k}\right)+1\right)^2 \left(16 \cos \left(\ft{4 \pi }{k}\right)+9 \cos \left(\ft{8 \pi }{k}\right)+21\right)\blue{\Q^4} +\mathcal O\left(\blue{\Q^5} \right),
\\[2mm]\notag
Y_{\rm w}{}&=    
   -\cot \left(\ft{2 \pi
   }{k}\right)-\cot \left(\ft{4 \pi }{k}\right)-\cot \left(\ft{6 \pi }{k}\right)-\cot
   \left(\ft{8 \pi }{k}\right)-\cot
   \left(\ft{10 \pi }{k}\right) 
\\[2mm]{}&    \notag
   -\left(32 \cos \left(\ft{4 \pi }{k}\right)+26 \cos \left(\ft{8 \pi }{k}\right)+20 \cos \left(\ft{12 \pi }{k}\right)+12 \cos \left(\ft{16 \pi }{k}\right)+6 \cos \left(\ft{20 \pi }{k}\right)+2 \cos \left(\ft{24 \pi }{k}\right)+17\right) \csc \left(\ft{8 \pi }{k}\right) \blue{\Q}
\\[2mm]{}&    \notag
   -\ft{1}{2}\big(333+644 \cos \left(\ft{4 \pi }{k}\right)+574 \cos \left(\ft{8 \pi }{k}\right)+472 \cos \left(\ft{12 \pi }{k}\right)+356 \cos \left(\ft{16 \pi }{k}\right)+248 \cos \left(\ft{20 \pi }{k}\right)+154 \cos \left(\ft{24 \pi }{k}\right)
\\[2mm]{}&   \notag
   +84 \cos \left(\ft{28 \pi }{k}\right)+36 \cos \left(\ft{32 \pi }{k}\right)+12 \cos \left(\ft{36 \pi }{k}\right)+2 \cos \left(\ft{40 \pi }{k}\right)\big)\csc \left(\ft{6 \pi }{k}\right)\sec \left(\ft{2 \pi }{k}\right)
   \blue{\Q^2} 
\\[2mm]{}&   \notag
   -\ft{1}{4} \big(19164 \cos \left(\ft{4 \pi }{k}\right)+17734 \cos \left(\ft{8 \pi }{k}\right)+15568 \cos \left(\ft{12 \pi }{k}\right)+12936 \cos \left(\ft{16 \pi }{k}\right)+10152 \cos \left(\ft{20 \pi }{k}\right)
\\[2mm]{}&   \notag
   +7494 \cos \left(\ft{24 \pi }{k}\right)+5186 \cos \left(\ft{28 \pi }{k}\right)+3336 \cos \left(\ft{32 \pi }{k}\right)+1980 \cos \left(\ft{36 \pi }{k}\right)+1062 \cos \left(\ft{40 \pi }{k}\right)+504 \cos \left(\ft{44 \pi }{k}\right)
\\[2mm]{}&   \notag   
   +204 \cos \left(\ft{48 \pi }{k}\right)+66 \cos \left(\ft{52 \pi }{k}\right)+16 \cos \left(\ft{56 \pi }{k}\right)+2 \cos \left(\ft{60 \pi }{k}\right)+9831\big) \csc \left(\ft{6 \pi }{k}\right) \sec \left(\ft{2 \pi }{k}\right) \sec \left(\ft{4 \pi }{k}\right)\blue{\Q^3}
\\[2mm]{}&    \notag
    -\ft{1}{4} \big(309516 \cos \left(\ft{4 \pi }{k}\right)+291096 \cos \left(\ft{8 \pi }{k}\right)+262674 \cos \left(\ft{12 \pi }{k}\right)+227222 \cos \left(\ft{16 \pi }{k}\right)+188204 \cos \left(\ft{20 \pi }{k}\right)
\\[2mm]{}&    \notag
    +148994 \cos \left(\ft{24 \pi }{k}\right)+112482 \cos \left(\ft{28 \pi }{k}\right)+80736 \cos \left(\ft{32 \pi }{k}\right)+54896 \cos \left(\ft{36 \pi }{k}\right)+35190 \cos \left(\ft{40 \pi }{k}\right)
\\[2mm]{}&    \notag
    +21134 \cos \left(\ft{44 \pi }{k}\right)+11774 \cos \left(\ft{48 \pi }{k}\right)+6010 \cos \left(\ft{52 \pi }{k}\right)+2754 \cos \left(\ft{56 \pi }{k}\right)+1108 \cos \left(\ft{60 \pi }{k}\right)+376 \cos \left(\ft{64 \pi }{k}\right)
\\[2mm]{}&    \notag
    +102 \cos \left(\ft{68 \pi }{k}\right)+20 \cos \left(\ft{72 \pi }{k}\right)+2 \cos \left(\ft{76 \pi }{k}\right)+157945\big) \csc \left(\ft{6 \pi }{k}\right) \sec \left(\ft{2 \pi }{k}\right) \sec \left(\ft{4 \pi }{k}\right)\blue{\Q^4}+O\left(\blue{\Q^5}\right) \ .
\end{align}

\bibliographystyle{JHEP}    
\bibliography{BKT2} 

\end{document}